# Estimating the multivariate extremal index function

CHRISTIAN Y. ROBERT

*ENSAE, Timbre J120, 3 avenue Pierre Larousse, 92245 Malakoff Cedex, France.*
*E-mail: chrobert@ensae.fr*

The multivariate extremal index function relates the asymptotic distribution of the vector of pointwise maxima of a multivariate stationary sequence to that of the independent sequence from the same stationary distribution. It also measures the degree of clustering of extremes in the multivariate process. In this paper, we construct nonparametric estimators of this function and prove their asymptotic normality under long-range dependence and moment conditions. The results are illustrated by means of a simulation study.

## 1. Introduction

The motivation for this paper comes from an empirical observation that time series from hydrology, meteorology, environmental sciences, finance, etc. are heavy-tailed and clustered when extremal events occur. In particular, it has been recognized in recent decades that the model of independent and identically distributed (i.i.d.) Gaussian random variables is inappropriate for modeling extreme returns of risky assets that are observed during a financial crisis. It is important for risk managers to understand the relative behavior of the various financial risks to which their institutions are exposed in the event of large losses because they have to anticipate the diversification opportunities so that the risks can be balanced by comovements (between risks) or reversal movements in short time intervals (within risks).

Although there are well-developed statistical approaches to characterize the cross-sectional dependence structure of extreme returns of risky assets (see, e.g., [14, 20, 23, 34] and the references therein), problems concerning the estimation of their temporal dependence structure have not received much attention. A notable exception is [46], which proposes a specific class of max-stable processes to model simultaneous dependencies between and within financial time series. However, this ad hoc class of processes is not necessarily suitable for any multivariate time series. The multivariate extremal index function, introduced by Nandagopalan [27, 28], is a quantity which allows one to relate the asymptotic distribution of the vector of pointwise maxima of a stationary sequence to that of the independent sequence from the same stationary distribution. It also measures the degree of clustering of extremes in the multivariate process since it is equal to the reciprocal of the mean number of clustered extremal events. Therefore, it is a specific measure of the temporal dependence structure of the extreme values of the process.

It is the aim of this paper to present a general theory for the inference of this function. We extend the block declustering approach introduced in [37] to the case of multivariate stationary processes: we construct pointwise estimators and study their asymptotic properties. Three assumptions are made: (i) there exist moment restrictions on the amount of clustering of extremes; (ii) the number of two-level exceedances converges weakly – an assumption which will guarantee the existence of the asymptotic variance–covariance matrix of the estimators; (iii) a mixing condition weaker than strong mixing is supposed to hold. Under these assumptions, we prove the asymptotic normality of our estimators.

More formally, let $(\boldsymbol{X}_l = (X_{l,1}, \ldots, X_{l,d}))_{l \geq 1}$ be a strictly stationary sequence with stationary distribution function $F(\boldsymbol{x}) = P(X_{l,i} \leq x_i, i=1,\ldots,d)$, $\boldsymbol{x} = (x_1, \ldots, x_d) \in \mathbb{R}^d$, and univariate marginal distributions $F_i(x) = P(X_{l,i} \leq x)$, $i=1,\ldots,d$. We assume that there exists a family of normalizing sequences in $\mathbb{R}^d$, $(\boldsymbol{u}_n(\boldsymbol{\tau}) = (u_{n,1}(\tau_1), \ldots, u_{n,d}(\tau_d)))_{n \geq 1}$, $\boldsymbol{\tau} = (\tau_1, \ldots, \tau_d) \in (0, \infty)^d$, such that

$$\lim_{n \to \infty} n(1 - F_i(u_{n,i}(\tau))) = \tau \qquad \text{for } \tau > 0, i = 1, \ldots, d, \tag{1.1}$$

and, for some function $\tilde{H} : (0, \infty)^d \mapsto [0, 1]$,

$$\lim_{n \to \infty} n(1 - F(\boldsymbol{u}_n(\boldsymbol{\tau}))) = -\ln \tilde{H}(\boldsymbol{\tau}), \qquad \text{for } \boldsymbol{\tau} \in (0, \infty)^d. \tag{1.2}$$

A necessary and sufficient condition for the existence of a sequence $(u_{n,i}(\tau))_{n \geq 1}$ which satisfies (1.1) is that $\lim_{x \to x_{f,i}} \bar{F}_i(x)/\bar{F}_i(x-) = 1$, where $x_{f,i} = \sup\{u : F_i(u) < 1\}$ and $\bar{F}_i = 1 - F_i$ (see Theorem 1.7.13 in [22]). A natural choice for $u_{n,i}(\tau)$ is then given by $F_i^{\leftarrow}(1 - \tau/n)$, $\tau \in [0, n)$, where $F_i^{\leftarrow}$ is the generalised inverse of $F_i$, that is, $F_i^{\leftarrow}(\tau) = \inf\{x \in \mathbb{R} : F_i(x) \geq \tau\}$. This assumption is weaker than assuming that $F_i$ is in the domain of attraction of an extreme value distribution since the normalization is linear in this case. However, the function $\tilde{G}$ defined by $\tilde{G}(\boldsymbol{\tau}) = \tilde{H}(\tau_1^{-1}, \ldots, \tau_d^{-1})$ for $\boldsymbol{\tau} \in (0, \infty)^d$ must be a multivariate extreme value distribution regardless of whether the normalization is linear (see [33], Proposition 2.1). In particular, $\tilde{G}$ is a continuous distribution function with unit Fréchet margins. It is noteworthy that $-\ln \tilde{H}$ is a homogeneous function of degree 1, that is, $-\ln \tilde{H}(c\boldsymbol{\tau}) = -c \ln \tilde{H}(\boldsymbol{\tau})$ for all $c > 0$ and $\boldsymbol{\tau} \in (0, \infty)^d$. This function is sometimes called the *stable tail dependence function* of $F$.

Let $M_{n,i} = \max(X_{1,i}, \ldots, X_{n,i})$ be the maximum of the $i$th component and introduce the vector of pointwise maxima $\boldsymbol{M}_n = (M_{n,1}, \ldots, M_{n,d})$. If $(\boldsymbol{X}_n)_{n \geq 1}$ is a sequence of independent and identically distributed (i.i.d.) vectors of random variables (r.v.s), then (1.2) is equivalent to

$$P(\boldsymbol{M}_n \leq \boldsymbol{u}_n(\boldsymbol{\tau})) = P(M_{n,i} \leq u_{n,i}(\tau_i), i = 1, \ldots, d) \to \tilde{H}(\boldsymbol{\tau}), \qquad \text{as } n \to \infty.$$

This convergence can be extended to stationary sequences by assuming the long-range dependence $D(\boldsymbol{u}_n(\boldsymbol{\tau}))$-condition introduced in [19], which is a natural multivariate version of the well-known univariate $D(u_n(\tau))$-condition (see, e.g., [22], page 53). Let $\mathbb{S}_+ \equiv \{\boldsymbol{\tau} \in (0,\infty)^d : |\boldsymbol{\tau}| = 1\}$, where $|\boldsymbol{\tau}|^2 = \sum_{i=1}^d \tau_i^2$. If $D(\boldsymbol{u}_n(\boldsymbol{\tau}))$ holds for each $\boldsymbol{\tau} \in (0,\infty)^d$ and $P(\boldsymbol{M}_n \leq \boldsymbol{u}_n(\boldsymbol{\tau}_0))$ converges as $n \to \infty$ for each $\boldsymbol{\tau}_0$ in $\mathbb{S}_+$, then there exists a function $\theta : (0,\infty)^d \mapsto [0,1]$ such that (see Proposition 2.3 in [33])

$$\lim_{n \to \infty} P(\boldsymbol{M}_n \leq \boldsymbol{u}_n(\boldsymbol{\tau})) = \tilde{H}(\boldsymbol{\tau})^{\theta(\boldsymbol{\tau})} \qquad \text{for } \boldsymbol{\tau} \in (0,\infty)^d. \tag{1.3}$$

Let $H(\boldsymbol{\tau}) = \tilde{H}(\boldsymbol{\tau})^{\theta(\boldsymbol{\tau})}$. The function $G$ defined by $G(\boldsymbol{\tau}) = H(\tau_1^{-1}, \ldots, \tau_d^{-1})$ is also a multivariate extreme value distribution and has Fréchet marginals. The function

$$\theta(\boldsymbol{\tau}) = \frac{-\ln H(\boldsymbol{\tau})}{-\ln \tilde{H}(\boldsymbol{\tau})} \tag{1.4}$$

is referred to as the *multivariate extremal index function* of $(\boldsymbol{X}_n)_{n \geq 1}$.

The estimation of the function $-\ln \tilde{H}$ for sequences of i.i.d. vectors of r.v.s has been thoroughly investigated (see, e.g., [9, 10, 11, 16, 18]). But, extensions to stationary sequences are still at an early stage (see an example in [42]). The estimation of the multivariate extremal index function has been little investigated. Recently, some pointwise estimators have been proposed, but their asymptotic properties have not been studied. In [41], Smith and Weissman introduce the class of multivariate maxima of moving maxima ($M_4$) processes and establish that the multivariate extremal index function of a very wide class of processes may be approximated arbitrarily closely by one from a $M_4$ process. Nevertheless, the estimation of $\theta$ via a $M_4$ process is practically infeasible since it necessitates the estimation of an infinite number of parameters, except if additional approximations are made. Smith and Weissman also give a key characterization of $\theta(\boldsymbol{\tau})$ as the univariate extremal index of a sequence depending on the standardized Fréchet components (see Proposition 2.1 in [41] and Proposition 2.1 below). It follows that if one can transform the data to have unit Fréchet components, then $\theta(\boldsymbol{\tau})$ may be estimated by univariate methods such as those proposed in [13, 40, 45], or [37]. To evaluate the accuracy of this approach, a simulation study is conducted in [12] with Ferro and Segers estimators (see [13]).

In this paper, we introduce two new nonparametric estimators of the multivariate extremal index function. Its original contribution is to study the asymptotic properties of these estimators. The paper is organized as follows. In Section 2, we discuss two characterizations of the multivariate extremal index function and present some of its properties. In Section 3, we explain how we construct the estimators. Note that they are based on a block declustering scheme and are only determined by the block length, as in [37]. In Section 4, we present and discuss technical conditions. We then give the asymptotic distributions of the estimators. In Section 5, we investigate their finite-sample behaviors on simulated data. The proof of the asymptotic normality of the estimators is found in Section 6. Section 7 concludes.

## 2. The multivariate extremal index

The multivariate extremal index function defined by (1.4) may also be characterized by the asymptotic distribution of the following point process of exceedances:

$$N_n^{(\boldsymbol{\tau})}(B) = \sum_{l=1}^{n} 1_{\{l/n \in B, \boldsymbol{X}_l \not\leq \boldsymbol{u}_n(\boldsymbol{\tau})\}},$$

where $B$ is a Borel set included in $(0,1]$ and $\{\boldsymbol{X}_l \not\leq \boldsymbol{u}_n(\boldsymbol{\tau})\} = \bigcup_{i=1}^{d} \{X_{l,i} > u_{n,i}(\tau_i)\}$. Contrary to the univariate case, there are several ways to define a multivariate threshold exceedance. Here, we define an exceedance as the event that one of the components of $\boldsymbol{X}_l$ exceeds its associated threshold. Suppose that (1.3) holds and $(\boldsymbol{X}_n)_{n \geq 1}$ satisfies the long-range dependence $\Delta(\boldsymbol{u}_n(\boldsymbol{\tau}))$-condition introduced in [28] (which is a little stronger than the $D(\boldsymbol{u}_n(\boldsymbol{\tau}))$-condition). A necessary and sufficient condition for the weak convergence of $N_n^{(\boldsymbol{\tau})}(\cdot)$ is then the convergence of $N_n^{(\boldsymbol{\tau})}((0; q_n/n])$ to a discrete distribution, $\pi^{(\boldsymbol{\tau})}$, given that there is at least one exceedance, that is,

$$\lim_{n \to \infty} P(N_n^{(\boldsymbol{\tau})}((0; q_n/n]) = k | N_n^{(\boldsymbol{\tau})}((0; q_n/n]) > 0) = \pi^{(\boldsymbol{\tau})}(k),$$

where $(q_n)$ is a $\Delta(\boldsymbol{u}_n(\boldsymbol{\tau}))$-separating sequence (see Section 4). $\pi^{(\boldsymbol{\tau})}$ is referred to as the *cluster-size distribution*. Under these assumptions, the point process converges to an homogeneous compound Poisson process, $N^{(\boldsymbol{\tau})}$, with intensity $-\theta(\boldsymbol{\tau})\ln \tilde{H}(\boldsymbol{\tau})$ and limiting compound distribution $\pi^{(\boldsymbol{\tau})}$. It may be stressed that, under some mild additional assumptions, we have the following characterization of the multivariate extremal index function (see [19, 24]):

$$\theta(\boldsymbol{\tau}) = \left(\sum_{k=1}^{\infty} k \pi^{(\boldsymbol{\tau})}(k)\right)^{-1},$$

that is, $\theta$ is equal to the reciprocal of the limiting mean number of exceedances in a cluster.

As mentioned in the Introduction, Smith and Weissman give an alternative characterization of the multivariate extremal index function in [41]. They first propose to standardize the margins to the unit Fréchet distribution and then to express $\theta(\boldsymbol{\tau})$ as the univariate extremal index of the constructed sequence as a linear combination of the standardized components. In this paper, we decide to standardize to the unit Pareto distribution, as in Section 10.5.2 of [4] (see property (v)).

**Proposition 2.1.** *Let $\boldsymbol{\tau} \in (0, \infty)^d \setminus \{\boldsymbol{0}\}$ and assume that $\lim_{n \to \infty} P(\boldsymbol{M}_n \leq \boldsymbol{u}_n(\boldsymbol{\tau})) = \tilde{H}(\boldsymbol{\tau})^{\theta(\boldsymbol{\tau})}$. Define the associated univariate stationary sequence by*

$$Z_l^{(\boldsymbol{\tau})} = \max_{i=1,\ldots,d} \tau_i Y_{l,i}, \qquad l \geq 1,$$

where $Y_{l,i} = (1 - F_{i,-}(X_{l,i}))^{-1}$ and $F_{i,-}(x) = P(X_{l,i} < x)$. $\theta(\boldsymbol{\tau})$ is then the univariate extremal index of the sequence $(Z_l^{(\boldsymbol{\tau})})_{l \geq 1}$, that is, it satisfies, for $\varkappa > 0$,

$$\lim_{n \to \infty} nP(Z_l^{(\boldsymbol{\tau})} > v_n^{(\boldsymbol{\tau})}(\varkappa)) = \varkappa \quad and \quad \lim_{n \to \infty} P\left(\max_{l=1,\ldots,n} Z_l^{(\boldsymbol{\tau})} \leq v_n^{(\boldsymbol{\tau})}(\varkappa)\right) = \mathrm{e}^{-\theta(\boldsymbol{\tau})\varkappa},$$

where $v_n^{(\boldsymbol{\tau})}(\varkappa) = \varkappa^{-1}(-\ln \tilde{H}(\boldsymbol{\tau}))n$.

The proof of Proposition 2.1 is postponed to Appendix A. Note that it completes the arguments introduced in Section 10.5.2 of [4], where it is assumed that the $F_i$ are continuous.

It is well known that, in the univariate case, $\theta$ is a constant which does not depend on $\tau$. In the multivariate case, $\theta$ is required to be a constant on the lines through the origin. In the next section, we will take into account this homogeneity property when constructing the estimators. More particularly, the multivariate extremal index function has the following properties (see [28, 33], Section 10.5.2 in [4], [24] and [25]):

1. $0 \leq \theta(\boldsymbol{\tau}) \leq 1$ for all $\boldsymbol{\tau} \in (0, \infty)^d$.
2. $\theta(\boldsymbol{\tau})$ is a continuous function of $\boldsymbol{\tau} \in (0, \infty)^d$ and is scale invariant, that is, $\theta(c\boldsymbol{\tau}) = \theta(\boldsymbol{\tau})$ for all $c > 0$ and $\boldsymbol{\tau} \in (0, \infty)^d$.
3. $\theta$ can be extended by continuity to $[0, \infty)^d \setminus \{\boldsymbol{0}\}$. Let $\boldsymbol{\tau}^{(i)} = (0, \ldots, 0, \tau_i, 0, \ldots, 0)$. The univariate extremal index, $\theta_i$, say, for the $i$th component sequence $(X_{n,i})_{n \geq 1}$ exists and $\theta_i = \theta(\boldsymbol{\tau}^{(i)})$. $\{\boldsymbol{0}\}$ is a discontinuity of $\theta$ if there exist $i$ and $j$ such that $\theta_i \neq \theta_j$. Note that the functions $H$ and $\tilde{H}$ can be extended by continuity to $[0, \infty)^d$. In particular, we have $\tilde{H}(\boldsymbol{\tau}^{(i)}) = \mathrm{e}^{-\tau_i}$, $H(\boldsymbol{\tau}^{(i)}) = \mathrm{e}^{-\theta_i \tau_i}$ and $\tilde{H}(\boldsymbol{0}) = H(\boldsymbol{0}) = 1$.
4. Bounds for $\theta(\boldsymbol{\tau})$ are given by

$$\frac{\max(\theta_1 \tau_1, \ldots, \theta_d \tau_d)}{-\ln \tilde{H}(\boldsymbol{\tau})} \leq \theta(\boldsymbol{\tau}) \leq \frac{\theta_1 \tau_1 + \cdots + \theta_d \tau_d}{-\ln \tilde{H}(\boldsymbol{\tau})}.$$

   The upper bound corresponds to the case where $G$ has independent components and the lower bound corresponds to the case where $G$ has totally dependent components.
5. If $G$ and $\tilde{G}$ have independent components, then $\theta(\boldsymbol{\tau}) = \sum_{i=1}^d \theta_i \tau_i / \sum_{i=1}^d \tau_i$.

In the next sections, we will illustrate our technical conditions and our limiting results with three examples of bivariate processes. Let us now introduce these processes and discuss their extremal properties. The first process will be considered as the benchmark because all the technical conditions can be easily verified and the calculations of the asymptotic variances of the estimators can be carried out explicitly. It is the bivariate process with independent univariate sequences and independent components. The second process is a bivariate squared ARCH(1) process with independent components. There is no cross-sectional dependence, but each component is time dependent. The third process is a bivariate autoregressive process of order 1 with dependent innovations. By modifying the values of the parameters of this process, we may have cross-sectional dependence or independence and temporal dependence or independence. Recall that for $d=2$, we have $(\boldsymbol{X}_l = (X_{l,1}, X_{l,2}))_{l \geq 1}$, $\boldsymbol{\tau} = (\tau_1, \tau_2)$ and $\boldsymbol{u}_n(\boldsymbol{\tau}) = (u_{n,1}(\tau_1), u_{n,2}(\tau_2))$.

***Example 2.1.*** The bivariate independent process with independent components: $X_{l,1} = \xi_{l,1}$ and $X_{l,2} = \xi_{l,2}$, where $(\xi_{l,1})_{l \geq 1}$ and $(\xi_{l,2})_{l \geq 1}$, are two independent sequences of i.i.d. standard exponential r.v.s. It is easily seen that $u_{n,1}(\tau) = u_{n,2}(\tau) = \ln(n/\tau)$, $-\ln \tilde{H}(\boldsymbol{\tau}) = \tau_1 + \tau_2$ and $\theta(\boldsymbol{\tau}) = 1$.

The cluster of exceedances of $N^{(\boldsymbol{\tau})}$ are of size 1, that is, the cluster-size distribution is given by $\pi^{(\boldsymbol{\tau})}(1) = 1$, $\pi^{(\boldsymbol{\tau})}(k) = 0$ for $k > 1$.

The associated series is given by $Z_l^{(\boldsymbol{\tau})} = \max(\tau_1 \exp(\xi_{l,1}), \tau_2 \exp(\xi_{l,2}))$ and we have $v_n^{(\boldsymbol{\tau})}(\varkappa) \sim n\varkappa^{-1}(\tau_1 + \tau_2)$ as $n \to \infty$.

***Example 2.2.*** A bivariate squared ARCH(1) process with independent components: $X_{l+1,i} = (\eta_i + \lambda_i X_{l,i}) \xi_{l+1,i}^2$ for $l \geq 1$ and $i = 1, 2$, where $(\xi_{l,1})_{l \geq 1}$ and $(\xi_{l,1})_{l \geq 2}$ are two independent sequences of i.i.d. standard Gaussian r.v.s, $\eta_i > 0$ and $0 < \lambda_i < 2e^{\gamma}$, where $\gamma$ is Euler's constant. We assume that $X_{1,1}$ and $X_{1,2}$ are drawn from the univariate stationary distributions. Let $\kappa_i$ be such that $\mathbb{E}(\lambda_i \xi_{l,i}^2)^{\kappa_i} = 1$ for $i = 1, 2$. There exist constants $c_i$ such that $\bar{F}_i(x) \sim c_i x^{-\kappa_i}$ as $x \to \infty$. It follows that $u_{n,i}(\tau) \sim (nc_i/\tau)^{1/\kappa_i}$ as $n \to \infty$ (see, e.g., [21] and [17]). Let $R_i(x) = \sharp\{j \geq 1 : \tilde{X}_i \prod_{l=1}^{j} (\lambda_i \xi_{l,i}^2) > x\}$ where $\tilde{X}_i$ is independent of $(\xi_{l,i})_{l \geq 1}$ and $P(\tilde{X}_i > x) = x^{-\kappa_i}$, $x \geq 1$, and define $p_{k,i} = P(R_i(1) = k)$, $k \geq 0$.

Since the components are independent, we have $-\ln \tilde{H}(\boldsymbol{\tau}) = \tau_1 + \tau_2$ and

$$\theta(\boldsymbol{\tau}) = \frac{\theta_1 \tau_1 + \theta_2 \tau_2}{\tau_1 + \tau_2}.$$

Moreover, $\theta_i = p_{0,i} = \int_1^{\infty} P(\bigvee_{j=1}^{\infty} \prod_{l=1}^{j} (\lambda_i \xi_{l,i}^2) \leq x) \kappa_i x^{-\kappa_i - 1} \, dx$, $i = 1, 2$.

The clusters of exceedances may be of any size. One can show that the cluster-size distribution of $N^{(\boldsymbol{\tau})}$ is given by

$$\pi^{(\boldsymbol{\tau})} = \frac{\theta_1 \tau_1}{\theta_1 \tau_1 + \theta_2 \tau_2} \pi_1 + \frac{\theta_2 \tau_2}{\theta_1 \tau_1 + \theta_2 \tau_2} \pi_2,$$

where $\pi_i(k) = (p_{k-1,i} - p_{k,i})/p_{0,i}$, $k \geq 1$ and $i = 1, 2$ (see [17]).

Since there is no analytic expression for the stationary univariate distributions, an explicit form of the associated sequence cannot be given.

***Example 2.3.*** A bivariate autoregressive process of order 1 with dependent innovations: $X_{l+1,i} = \rho_i X_{l,i} + \xi_{l+1,i}$ for $l \geq 1$ and $i = 1, 2$, where $(\xi_{l,1}, \xi_{l,2})_{l \geq 1}$ is a sequence of i.i.d. vectors with a bivariate unit Fréchet extreme value distribution, that is,

$$P(\xi_{l,1} \le x_1, \xi_{l,2} \le x_2) = \exp\left(-\left(\frac{1}{x_1} + \frac{1}{x_2}\right) A\left(\frac{x_1}{x_1+x_2}\right)\right)$$

$$:= \exp(-B(x_1, x_2)),$$

where $A$ is a convex and differentiable function bounded below by $\max(x, 1-x)$ and above by 1. We assume that $0 < \rho_i < 1$ and that $(X_{1,1}, X_{1,2})$ is drawn from the stationary distribution. We have that $\bar{F}_i(x) \sim (1-\rho_i)^{-1}x^{-1}$ as $x \to \infty$ and it follows that $u_{n,i}(\tau) \sim n/((1-\rho_i)\tau)$ as $n \to \infty$. By Theorem 2.1 in [36], we deduce that

$$-\ln \tilde{H}(\boldsymbol{\tau}) = \sum_{k=0}^{\infty} B(((1-\rho_1)\rho_1^k \tau_1)^{-1}, ((1-\rho_2)\rho_2^k \tau_2)^{-1}).$$

Similar arguments as in Section 6 of [33] show that

$$\theta(\boldsymbol{\tau}) = \frac{-\ln \tilde{H}((\tau_1, \tau_2)) + \ln \tilde{H}((\rho_1 \tau_1, \rho_2 \tau_2))}{-\ln \tilde{H}((\tau_1, \tau_2))}.$$

It is important to note that if $\rho_1 = \rho_2 = \rho$, then $-\ln \tilde{H}(\boldsymbol{\tau}) = B(\tau_1^{-1}, \tau_2^{-1})$ and $\theta(\boldsymbol{\tau}) = (1-\rho)$. The multivariate index does not depend on $\boldsymbol{\tau}$. If $A = 1$, that is, if $\xi_{l,1}$ and $\xi_{l,2}$ are independent, then $-\ln \tilde{H}(\boldsymbol{\tau}) = \tau_1 + \tau_2$ and $\theta(\boldsymbol{\tau}) = ((1-\rho_1)\tau_1 + (1-\rho_2)\tau_2)/(\tau_1 + \tau_2)$.

The clusters of exceedances may be of any size. The asymptotic distribution of $N_n^{(\boldsymbol{\tau})}$ may be obtained by using results of Section 2 in [6]. Moreover, since there is no analytic expression for the stationary bivariate distribution, an explicit form of the associated sequence cannot be given.

## 3. Defining the estimators

In this section, we explain our approach to estimating the multivariate extremal index function. As in [37], we consider a block declustering scheme and estimate intermediate thresholds such that we only have to take into account the block length to study the asymptotic distribution of the estimators.

Let us divide $[1, \ldots, n]$ into $k_n$ blocks of length $r_n$ ($k_n$ is the integer part of $n/r_n$), $I_j = [(j-1)r_n + 1, \ldots, jr_n]$ for $j = 1, \ldots, k_n$ and a last block $I_{k_n+1} = [r_n k_n + 1, \ldots, n]$. The number of exceedances for the $j$th block is defined by $N_{r_n,j}^{(\boldsymbol{\tau})} = \sum_{l \in I_j} 1_{\{\boldsymbol{X}_l \not\le \boldsymbol{u}_{r_n}(\boldsymbol{\tau})\}}$ for $j = 1, \ldots, k_n$, where $u_{r_n,i}(\tau_i) = F_i^{\leftarrow}(1 - \tau_i/r_n)$, $i = 1, \ldots, d$. The main issue when using these quantities to construct estimators is that the thresholds $u_{r_n,i}(\tau_i)$ are unknown since they depend on the univariate marginals of the stationary distribution. They have to be estimated from the data. As in [37], we consider estimators of the thresholds which are based on the order statistics. If $0 < \tau \le r_n$, let $\hat{u}_{r_n,i}(\tau) = X_{(\lceil k_n \tau \rceil), i}$, where $X_{(k), i}$ is the $k$th largest of $X_{1,i}, \ldots, X_{k_n r_n, i}$ and $\lceil x \rceil$ denotes the smallest integer greater than or equal to $x$. If $\tau = 0$, let $\hat{u}_{r_n,i}(0) = \infty$. Now, define $\hat{N}_{r_n,j}^{(\tau)} = \sum_{l \in I_j} 1_{\{X_l \nleq \hat{u}_{r_n}(\tau)\}}$ for $\tau \in [0, r_n]^d$.

In order to estimate the multivariate extremal index function, it seems natural to exploit the characterization given by (1.4). Let $N_n^{(\tau)} \equiv N_n^{(\tau)}((0,1])$ and note that, under appropriate conditions (see the following section),

$$\lim_{n \to \infty} P(N_n^{(\tau)} = 0) = \lim_{n \to \infty} P(\boldsymbol{M}_n \leq \boldsymbol{u}_n(\tau)) = H(\tau)$$

and that, by (1.2),

$$\lim_{n \to \infty} \mathbb{E}(N_n^{(\tau)}) = \lim_{n \to \infty} n(1 - F(\boldsymbol{u}_n(\tau))) = -\ln \tilde{H}(\tau).$$

Let us use the empirical distribution of the number of exceedances to provide empirical counterparts of $H(\tau)$ and $-\ln \tilde{H}(\tau)$. We define

$$\hat{H}_n(\tau) = \frac{1}{k_n} \sum_{j=1}^{k_n} 1_{\{\hat{N}_{r_n,j}^{(\tau)} = 0\}} \quad \text{and} \quad -\ln \widehat{\tilde{H}}_n(\tau) = \frac{1}{k_n} \sum_{j=1}^{k_n} \hat{N}_{r_n,j}^{(\tau)} \quad \text{for } \tau \in [0, r_n]^d.$$

One may consider $-\ln \hat{H}_n(\tau)/(-\ln \widehat{\tilde{H}}_n(\tau))$ in order to estimate $\theta(\tau)$. But, unlike the multivariate extremal function, this function is not scale invariant (see Property 2 in the previous section). Hence, we introduce a first estimator which satisfies the homogeneity property:

$$\hat{\theta}_n^{(1)}(\tau) = \frac{-\ln \hat{H}_n(\tau/L(\tau))}{-\ln \widehat{\tilde{H}}_n(\tau/L(\tau))}, \qquad \tau/L(\tau) \in [0, r_n]^d \setminus \{\boldsymbol{0}\},$$

where $L$ is a known function from $[0, \infty)^d \setminus \{\boldsymbol{0}\}$ to $(0, \infty)$ which is homogeneous of order 1. For example, consider the family $L_{c,a}(\tau) = c(\sum_{i=1}^d |\tau_i|^a)^{1/a}$ for $a > 0$ and $c > 0$.

The second estimator is derived from the characterization of Proposition 2.1. Let us consider the number of exceedances of $(Z_n^{(\tau)})_{n \geq 1}$ above the threshold $v_n^{(\tau)}(\varkappa)$:

$$N_n^{(\varkappa, \tau)} = \sum_{l=1}^n 1_{\{Z_l^{(\tau)} > v_n^{(\tau)}(\varkappa)\}}.$$

Proposition 2.1 implies that $\lim_{n \to \infty} -\ln P(N_n^{(\varkappa, \tau)} = 0) = \theta(\tau)\varkappa$. In order to construct an alternative estimator of the extremal index function, we can follow the approach developed in [37]. First, we replace the $Z_l^{(\tau)}$ by their empirical counterparts since the marginal distribution functions $F_i$ are unknown. Let $R_{l,i}$ denote the rank of $X_{l,i}$ among $(X_{1,i}, \ldots, X_{k_n r_n, i})$. In the case of ties, the lowest rank for the ties is used for each tie. We define

$$\check{Z}_l^{(\boldsymbol{\tau})} = \max_{i=1,\ldots,d} \tau_i \check{Y}_{l,i},$$

where

$$\check{Y}_{l,i} = \frac{k_n r_n}{k_n r_n + 1 - R_{l,i}}.$$

We then introduce the number of exceedances of $(\check{Z}_l^{(\boldsymbol{\tau})})_{l \in I_j}$ for the $j$th block: $N_{r_n,j}^{(\varkappa,\boldsymbol{\tau})} = \sum_{l \in I_j} 1_{\{\check{Z}_l^{(\boldsymbol{\tau})} > v_{r_n}^{(\boldsymbol{\tau})}(\varkappa)\}}$. As previously, $v_{r_n}^{(\boldsymbol{\tau})}(\varkappa)$ is unknown. However, it may be estimated by $\hat{v}_{r_n}^{(\boldsymbol{\tau})}(\varkappa) = \check{Z}_{(\lceil k_n \varkappa \rceil)}^{(\boldsymbol{\tau})}$, where $\check{Z}_{(\lceil k_n \varkappa \rceil)}^{(\boldsymbol{\tau})}$ is the $(\lceil k_n \varkappa \rceil)$th-largest value among $\check{Z}_1^{(\boldsymbol{\tau})}, \ldots, \check{Z}_{k_n r_n}^{(\boldsymbol{\tau})}$. Finally, let us define $\hat{N}_{r_n,j}^{(\varkappa,\boldsymbol{\tau})}$ as the counterpart of $N_{r_n,j}^{(\varkappa,\boldsymbol{\tau})}$, where $v_{r_n}^{(\boldsymbol{\tau})}(\varkappa)$ is replaced by $\hat{v}_{r_n}^{(\boldsymbol{\tau})}(\varkappa)$, and introduce the second estimator

$$\hat{\theta}_n^{(2)}(\boldsymbol{\tau}) = -\varkappa^{-1} \ln\left(\frac{1}{k_n} \sum_{j=1}^{k_n} 1_{\{\hat{N}_{r_n,j}^{(\varkappa,\boldsymbol{\tau})} = 0\}}\right), \qquad \boldsymbol{\tau} \in [0,\infty)^d \setminus \{\mathbf{0}\}, \varkappa \in (0, r_n].$$

Note that this estimator is scale invariant without transformation on $\boldsymbol{\tau}$.

***Remark 3.1.*** In [37], three estimators of the univariate extremal index are introduced. The first estimator, denoted by $\hat{\theta}_{1,n}^{(\cdot)}$, is very close of our estimators $\hat{\theta}_n^{(1)}$ and $\hat{\theta}_n^{(2)}$ when they are evaluated at the points $\boldsymbol{\tau} = \boldsymbol{\tau}^{(i)}$, $i = 1, \ldots, d$. In fact, if $L$ is assumed to be a constant equal to 1 and $\varkappa = \tau_i$, we have

$$\hat{\theta}_n^{(1)}(\boldsymbol{\tau}^{(i)}) = \hat{\theta}_{1,n}^{(\tau_i)} \frac{k_n \tau_i}{\lceil k_n \tau_i \rceil - 1} \quad \text{and} \quad \hat{\theta}_n^{(2)}(\boldsymbol{\tau}^{(i)}) = \hat{\theta}_{1,n}^{(\tau_i)}.$$

It follows that in the univariate case (i.e., $d = 1$), both estimators have the same asymptotic behavior as $\hat{\theta}_{1,n}^{(\cdot)}$.

## 4. Main result

In this section, we first present and discuss technical conditions which are required for the asymptotic normality of the estimators. These conditions are quite similar to conditions introduced in [37] which are used, in particular, to establish the asymptotic properties of the estimator of the univariate extremal index $\hat{\theta}_{1,n}^{(\tau)}$ (see Remark 3.1 above). They might appear quite stringent in comparison with those of [45], where the asymptotic properties of the blocks and runs estimators of the univariate extremal index are studied. This is not the case for two reasons. First, we estimate intermediate thresholds and do not consider them as tuning parameters, contrary to [45]. This allows us to establish the asymptotic properties of intermediate empirical processes, which is more complicated and necessitates more conditions. Second, these conditions guarantee the existence of the asymptotic variances of the estimators, whereas [45] just assumes the convergence of the variance of a partial sum to the asymptotic variance and does not give any condition such that this convergence holds. Finally, one can refer to Section 4 of [37] for a comparison of similar conditions to those in [38] that are needed for convergence of the tail empirical process of a univariate stationary sequence.

Let us turn to some definitions which are the natural multivariate versions of definitions from [32] (see also [28] and [33]).

**Definition 4.1.** *Fix an integer $m \geq 1$. Let $\mathcal{F}_{p,q} = \mathcal{F}_{p,q}(\boldsymbol{\tau}_1, \ldots, \boldsymbol{\tau}_m)$ be the $\sigma$-algebra generated by the events $\{\boldsymbol{X}_l \not\leq \boldsymbol{u}_n(\boldsymbol{\tau}_j)\}$, $p \leq l \leq q$ and $1 \leq j \leq m$, and let*

$$\alpha_{n,l}(\boldsymbol{\tau}_1, \ldots, \boldsymbol{\tau}_m) \equiv \sup |P(A \cap B) - P(A)P(B) : A \in \mathcal{F}_{1,t}, B \in \mathcal{F}_{t+l,n}, 1 \leq t \leq n-l|.$$

*The $\Delta(\{\boldsymbol{u}_n(\boldsymbol{\tau}_j)\}_{1 \leq j \leq m})$-condition is said to hold if $\lim_{n \to \infty} \alpha_{n,l_n}(\boldsymbol{\tau}_1, \ldots, \boldsymbol{\tau}_m) = 0$ for some sequence $l_n = o(n)$.*

**Definition 4.2.** *Suppose that the $\Delta(\{\boldsymbol{u}_n(\boldsymbol{\tau}_j)\}_{1 \leq j \leq m})$-condition holds. A sequence of positive integers $(q_n)_{n \geq 1}$ is said to be $\Delta(\{\boldsymbol{u}_n(\boldsymbol{\tau}_j)\}_{1 \leq j \leq m})$-separating if, as $n \to \infty$, $q_n = o(n)$ and there exists a sequence $(l_n)_{n \geq 1}$ such that $\lim_{n \to \infty} nq_n^{-1} \alpha_{n,l_n}(\boldsymbol{\tau}_1, \ldots, \boldsymbol{\tau}_m) = 0$ and $l_n = o(q_n)$.*

We now give a decomposition of the numbers of exceedances when considering two vectors of thresholds, $\boldsymbol{u}_n(\boldsymbol{\tau}_1)$ and $\boldsymbol{u}_n(\boldsymbol{\tau}_2)$ for $\boldsymbol{\tau}_1, \boldsymbol{\tau}_2 \in [0, \infty)^d$. We define

$$N_{n,0,p}^{(\boldsymbol{\tau}_1, \boldsymbol{\tau}_2)} = \sum_{l=1}^{p} 1_{\{\boldsymbol{X}_l \not\leq \boldsymbol{u}_n(\boldsymbol{\tau}_1)\} \cup \{\boldsymbol{X}_l \not\leq \boldsymbol{u}_n(\boldsymbol{\tau}_2)\}},$$

$$N_{n,1,p}^{(\boldsymbol{\tau}_1, \boldsymbol{\tau}_2)} = \sum_{l=1}^{p} 1_{\{\boldsymbol{X}_l \not\leq \boldsymbol{u}_n(\boldsymbol{\tau}_1)\} \setminus \{\boldsymbol{X}_l \not\leq \boldsymbol{u}_n(\boldsymbol{\tau}_2)\}},$$

$$N_{n,2,p}^{(\boldsymbol{\tau}_1, \boldsymbol{\tau}_2)} = \sum_{l=1}^{p} 1_{\{\boldsymbol{X}_l \not\leq \boldsymbol{u}_n(\boldsymbol{\tau}_2)\} \setminus \{\boldsymbol{X}_l \not\leq \boldsymbol{u}_n(\boldsymbol{\tau}_1)\}},$$

$$N_{n,3,p}^{(\boldsymbol{\tau}_1, \boldsymbol{\tau}_2)} = \sum_{l=1}^{p} 1_{\{\boldsymbol{X}_l \not\leq \boldsymbol{u}_n(\boldsymbol{\tau}_1)\} \cap \{\boldsymbol{X}_l \not\leq \boldsymbol{u}_n(\boldsymbol{\tau}_2)\}}.$$

Note that $N_{n,0,p}^{(\boldsymbol{\tau}_1, \boldsymbol{\tau}_2)} = \sum_{i=1}^{3} N_{n,i,p}^{(\boldsymbol{\tau}_1, \boldsymbol{\tau}_2)}$ and $N_n^{(\boldsymbol{\tau}_i)} = N_{n,i,n}^{(\boldsymbol{\tau}_1, \boldsymbol{\tau}_2)} + N_{n,3,n}^{(\boldsymbol{\tau}_1, \boldsymbol{\tau}_2)}$, $i = 1, 2$.

We continue by presenting the first technical condition and then discussing the weak convergence of the sequence $(N_{n,1,n}^{(\boldsymbol{\tau}_1, \boldsymbol{\tau}_2)}, N_{n,2,n}^{(\boldsymbol{\tau}_1, \boldsymbol{\tau}_2)}, N_{n,3,n}^{(\boldsymbol{\tau}_1, \boldsymbol{\tau}_2)})_{n \geq 1}$.

***Condition (*C1*).***

(i) *The stationary sequence $(\boldsymbol{X}_n)_{n\geq 1}$ has a multivariate extremal index function $\theta > 0$.*

(ii) *For each $\boldsymbol{\tau}_1, \boldsymbol{\tau}_2 \in [0,\infty)^d \backslash \{\boldsymbol{0}\}$, the $\Delta(\boldsymbol{u}_n(\boldsymbol{\tau}_1), \boldsymbol{u}_n(\boldsymbol{\tau}_2))$-condition holds and there exists a probability measure $\pi^{(\boldsymbol{\tau}_1, \boldsymbol{\tau}_2)}$ such that for all $i_1 \geq 0, i_2 \geq 0, i_3 \geq 0, i_1 + i_2 + i_3 \geq 1$,*

$$\pi^{(\boldsymbol{\tau}_1,\boldsymbol{\tau}_2)}(i_1,i_2,i_3) = \lim_{n\to\infty} P(N_{n,h,q_n}^{(\boldsymbol{\tau}_1,\boldsymbol{\tau}_2)} = i_h, h=1,2,3 | N_{n,0,q_n}^{(\boldsymbol{\tau}_1,\boldsymbol{\tau}_2)} > 0) \tag{C1.a}$$

*for some $\Delta(\boldsymbol{u}_n(\boldsymbol{\tau}_1), \boldsymbol{u}_n(\boldsymbol{\tau}_2))$-separating sequence $(q_n)_{n\geq 1}$.*

For $\boldsymbol{\tau}_i = (\tau_{1,i}, \ldots, \tau_{d,i})$ and $i=1,2$, let $\boldsymbol{\tau}_1 \vee \boldsymbol{\tau}_2 = (\tau_{1,1} \vee \tau_{1,2}, \ldots, \tau_{d,1} \vee \tau_{d,2})$. For each $\boldsymbol{\tau}_1, \boldsymbol{\tau}_2 \in [0,\infty)^d \backslash \{\boldsymbol{0}\}$, let $\boldsymbol{\zeta} \equiv (\zeta_{l,1}^{(\boldsymbol{\tau}_1,\boldsymbol{\tau}_2)}, \zeta_{l,2}^{(\boldsymbol{\tau}_1,\boldsymbol{\tau}_2)}, \zeta_{l,3}^{(\boldsymbol{\tau}_1,\boldsymbol{\tau}_2)})_{l \geq 1}$ be an sequence of i.i.d. vectors of integer r.v.s with distribution $\pi^{(\boldsymbol{\tau}_1,\boldsymbol{\tau}_2)}$ and $\eta(\boldsymbol{\tau}_1, \boldsymbol{\tau}_2)$ be an r.v. with Poisson distribution and parameter $-\theta(\boldsymbol{\tau}_1 \vee \boldsymbol{\tau}_2) \ln(\tilde{H}(\boldsymbol{\tau}_1 \vee \boldsymbol{\tau}_2))$ independent of the sequence $\boldsymbol{\zeta}$.

The probability measure $\pi^{(\boldsymbol{\tau}_1,\boldsymbol{\tau}_2)}$ is a key parameter to characterize the distribution of the limiting two-level exceedance point process (see Theorem 2.5 and its proof in [32] for the univariate case). The following proposition is concerned with the weak convergence of the related sequence of the numbers of exceedances.

**Proposition 4.1.** *Suppose that (C1) holds. Then,*

$$\begin{aligned}(N_{n,1,n}^{(\boldsymbol{\tau}_1,\boldsymbol{\tau}_2)}, N_{n,2,n}^{(\boldsymbol{\tau}_1,\boldsymbol{\tau}_2)}, N_{n,3,n}^{(\boldsymbol{\tau}_1,\boldsymbol{\tau}_2)}) &\xrightarrow{D} (N_1^{(\boldsymbol{\tau}_1,\boldsymbol{\tau}_2)}, N_2^{(\boldsymbol{\tau}_1,\boldsymbol{\tau}_2)}, N_3^{(\boldsymbol{\tau}_1,\boldsymbol{\tau}_2)}) \\ &\stackrel{D}{=} \sum_{l=1}^{\eta(\boldsymbol{\tau}_1,\boldsymbol{\tau}_2)} (\zeta_{l,1}^{(\boldsymbol{\tau}_1,\boldsymbol{\tau}_2)}, \zeta_{l,2}^{(\boldsymbol{\tau}_1,\boldsymbol{\tau}_2)}, \zeta_{l,3}^{(\boldsymbol{\tau}_1,\boldsymbol{\tau}_2)}).\end{aligned} \tag{4.1}$$

*Moreover $\pi^{(\boldsymbol{\tau}_1,\boldsymbol{\tau}_2)}$ is scale invariant, that is, for each $\boldsymbol{\tau}_1, \boldsymbol{\tau}_2 \in [0,\infty)^d \backslash \{\boldsymbol{0}\}$ and $c > 0$, $\pi^{(c\boldsymbol{\tau}_1, c\boldsymbol{\tau}_2)} = \pi^{(\boldsymbol{\tau}_1, \boldsymbol{\tau}_2)}$.*

The proof of Proposition 4.1 is postponed to Appendix A.

Let us examine the distribution of the cluster sizes of the two-level exceedances, $\pi^{(\boldsymbol{\tau}_1,\boldsymbol{\tau}_2)}$, for the first example introduced in Section 2. The distribution for the second example may be derived by considering its Laplace transform. The distribution for the third example may be derived by using results of Section 2 in [6].

*Example 2.1 (continued).* The clusters of the two-level exceedances are of size 0 or 1. More precisely, the distribution of the cluster sizes is given by

$$\pi^{(\boldsymbol{\tau}_1,\boldsymbol{\tau}_2)}(1,0,0) = \frac{(\tau_{2,1} - \tau_{2,2})^+ + (\tau_{1,1} - \tau_{1,2})^+}{\tau_{1,1} \vee \tau_{1,2} + \tau_{2,1} \vee \tau_{2,2}},$$

$$\pi^{(\boldsymbol{\tau}_1,\boldsymbol{\tau}_2)}(0,1,0) = \frac{(\tau_{2,2} - \tau_{2,1})^+ + (\tau_{1,2} - \tau_{1,1})^+}{\tau_{1,1} \vee \tau_{1,2} + \tau_{2,1} \vee \tau_{2,2}},$$

$$\pi^{(\boldsymbol{\tau}_1,\boldsymbol{\tau}_2)}(0,0,1) = 1 - \pi_2^{(\boldsymbol{\tau}_1,\boldsymbol{\tau}_2)}(1,0,0) - \pi_2^{(\boldsymbol{\tau}_1,\boldsymbol{\tau}_2)}(0,1,0).$$

Let us turn to the second technical condition which is a multivariate version of Condition (C2) in [37]. Note that, since the estimated thresholds for our estimators are contingent on $k_n \tau_i$, $i = 1, \ldots, d$, and $k_n \varkappa$, and since $k_n$ may be chosen up to a proportional factor, we can assume, without loss of generality, that $\boldsymbol{\tau}$ and $\varkappa$ are bounded. Hence, let us now assume that $\boldsymbol{\tau} \in [0,1]^d$.

*Condition (**C2**).*

(i) *Let $r > 4d$. There exists a constant $D = D(r) \geq 0$ such that for all $\boldsymbol{\tau}_1, \boldsymbol{\tau}_2 \in [0,1]^d$,*

$$\sup_{n \geq 1} \mathbb{E}|N_n^{(\boldsymbol{\tau}_1)} - N_n^{(\boldsymbol{\tau}_2)}|^r \leq D|\boldsymbol{\tau}_1 - \boldsymbol{\tau}_2|. \tag{C2.a}$$

(ii) *Let $\omega > (4d-1)r/(r-4d)$. There exists a constant $C \geq 0$ such that for every choice of $\boldsymbol{\tau}_1, \ldots, \boldsymbol{\tau}_m \in [0,1]^d$, $m \geq 1$ and $n \geq l \geq 1$,*

$$\alpha_{n,l}(\boldsymbol{\tau}_1, \ldots, \boldsymbol{\tau}_m) \leq \alpha_l := C l^{-\omega}. \tag{C2.b}$$

(iii) *$(r_n)_{n \geq 1}$ is a sequence such that $r_n \to \infty$ and $r_n = o(n)$ and there exists a sequence $(l_n)_{n \geq 1}$ satisfying*

$$l_n = o(r_n^{2/r}) \quad \text{and} \quad \lim_{n \to \infty} n r_n^{-1} \alpha_{l_n} = 0. \tag{C2.c}$$

Let us describe some intuitions regarding this condition. First, (C2)(i) restricts the size of clusters by assuming that $N_n^{(\boldsymbol{\tau}_1)} - N_n^{(\boldsymbol{\tau}_2)}$ has a suitably bounded $r$th moment. It provides an inequality which will very useful to prove tightness criteria for intermediate empirical processes introduced in Section 5. Note that

$$|N_n^{(\boldsymbol{\tau}_1)} - N_n^{(\boldsymbol{\tau}_2)}| \leq \sum_{i=1}^{d} (N_n^{(\boldsymbol{\tau}_1^{(i)} \wedge \boldsymbol{\tau}_2^{(i)})} - N_n^{(\boldsymbol{\tau}_1^{(i)} \vee \boldsymbol{\tau}_2^{(i)})}).$$

It follows that it is sufficient to show that for each $i = 1, \ldots, d$, there exists a constant $D_i \geq 0$ such that for all $\boldsymbol{\tau}_1, \boldsymbol{\tau}_2 \in [0,1]^d$,

$$\sup_{n \geq 1} \mathbb{E}(N_n^{(\boldsymbol{\tau}_1^{(i)} \wedge \boldsymbol{\tau}_2^{(i)})} - N_n^{(\boldsymbol{\tau}_1^{(i)} \vee \boldsymbol{\tau}_2^{(i)})})^r \leq D_i |\tau_{i,1} - \tau_{i,2}|. \tag{4.2}$$

It is easily seen that (C2)(ii) is satisfied by strongly mixing stationary sequences where the mixing coefficients vanish with at least a sufficient hyperbolic rate. The underlying idea of the block declustering scheme is to split the block $I_j$ into a small block of length $l_n$ and a large block of length $r_n - l_n$. (C2)(ii) and (C2)(iii) essentially means that $l_n$ is sufficiently large such that blocks that are not adjacent are asymptotically independent, but does not grow too fast so that the contributions of the small blocks is negligible. Let us give now some clues explaining why this condition holds for the examples introduced in Section 2.

***Example 2.1 (continued).*** $((X_{l,1}, X_{l,2}))_{l\geq 1}$ is an i.i.d. sequence. Therefore, $\alpha_{n,l}(\boldsymbol{\tau}_1, \ldots, \boldsymbol{\tau}_m) = 0$ for every choice of $\boldsymbol{\tau}_1, \ldots, \boldsymbol{\tau}_m \in [0,1]^d$, $m \geq 1$, $n \geq 1$, $l \geq 1$. Moreover, $N_n^{(\boldsymbol{\tau}_1^{(i)} \wedge \boldsymbol{\tau}_2^{(i)})} - N_n^{(\boldsymbol{\tau}_1^{(i)} \vee \boldsymbol{\tau}_2^{(i)})}$ has a binomial distribution with parameters $n$ and $|\tau_{i,1} - \tau_{i,2}|/n$. Condition (4.2) is easily verified for any integer $r$.

***Example 2.2 (continued).*** The components of $((X_{l,1}, X_{l,2}))_{l\geq 1}$ are independent and each component is geometrically strong-mixing (see Example 3.1 in [37]). Moreover, bounds for the moment condition (4.2) can be obtained for any integer $r$ by using the same arguments as for Lemma 6.1 in [37].

***Example 2.3 (continued).*** $((X_{l,1}, X_{l,2}))_{l\geq 1}$ is a bivariate positive Harris recurrent Markov chain. Moreover, it is a particular case of a first-order stochastic equations with random coefficients. Following [26], one can show that $((X_{l,1}, X_{l,2}))_{l\geq 1}$ is geometrically absolute regular and strong-mixing. Moreover, bounds for the moment condition (4.2) can be obtained for any integer $r$ by using the Markov property of the components and similar arguments as for Lemma 6.1 in [37].

Finally, we assume that the convergence rate of $r_n$ to infinity is such that the bias of our estimators is asymptotically negligible with respect to their variance. Moreover, we need a condition on the regularity of $H$ and $\tilde{H}$ to guarantee that the asymptotic distribution is Gaussian.

***Condition (C3).***

(i) *The sequence* $(r_n)_{n\geq 1}$ *satisfies*

$$\lim_{n\to\infty} \sqrt{k_n} \sup_{\boldsymbol{\tau}\in[0,1]^d} |P(N_{r_n,1}^{(\boldsymbol{\tau})} = 0) - H(\boldsymbol{\tau})| = 0$$

*and*

$$\lim_{n\to\infty} \sqrt{k_n} \sup_{\boldsymbol{\tau}\in[0,1]^d} |r_n(1 - F(\boldsymbol{u}_{r_n}(\boldsymbol{\tau}))) + \ln \tilde{H}(\boldsymbol{\tau})| = 0.$$

(ii) *The functions $H$ and $\tilde{H}$ are (Fréchet) differentiable on $(0,1)^d$ and their derivatives can be extended by continuity to $[0,1]^d$.*

***Example 2.1 (continued).*** Note that as $n \to \infty$,

$$\sqrt{k_n} \sup_{\boldsymbol{\tau}\in[0,1]^d} |P(N_{r_n,1}^{(\boldsymbol{\tau})} = 0) - H(\boldsymbol{\tau})| \sim \frac{\mathrm{e}^{-1}}{2} \frac{\sqrt{k_n}}{r_n} \sim \frac{\mathrm{e}^{-1}}{2} \frac{n^{1/2}}{r_n^{3/2}},$$

$$\sqrt{k_n}\sup_{\boldsymbol{\tau}\in[0,1]^d}|r_n(1-F(\boldsymbol{u}_{r_n}(\boldsymbol{\tau})))+\ln\tilde{H}(\boldsymbol{\tau})|=\frac{\sqrt{k_n}}{r_n}\sim\frac{n^{1/2}}{r_n^{3/2}}.$$

It follows that if $r_n = o(n^{1/3})$, then Condition (C3) does not hold.

We end this section by giving the distributional asymptotics of the estimators. Let $\Theta(\cdot)$ be a pathwise continuous Gaussian process on $[0,1]^d\setminus\{\boldsymbol{0}\}$ with covariance function given in Appendix B.

Let $\Psi_L \equiv \{\boldsymbol{\tau}:\boldsymbol{\tau}/L(\boldsymbol{\tau})\in[0,1]^d\setminus\{\boldsymbol{0}\}\}$ and $\Psi_\varkappa \equiv \{\boldsymbol{\tau}:\varkappa\boldsymbol{\tau}/(-\ln\tilde{H}(\boldsymbol{\tau}))\in[0,1]^d\setminus\{\boldsymbol{0}\}\}$.

**Theorem 4.1.** *Suppose that (C1), (C2) and (C3) hold. If we let $m \geq 1$ and $\boldsymbol{\tau}_1,\ldots,\boldsymbol{\tau}_m \in \Psi_L$, then*

$$\sqrt{k_n}(\hat{\theta}_n^{(1)}(\boldsymbol{\tau}_i)-\theta(\boldsymbol{\tau}_i))_{i=1,\ldots,m} \overset{D}{\to} (\Theta(\boldsymbol{\tau}_i L^{-1}(\boldsymbol{\tau}_i)))_{i=1,\ldots,m}.$$

*If we let $m \geq 1$ and $\boldsymbol{\tau}_1,\ldots,\boldsymbol{\tau}_m \in \Psi_\varkappa$, then*

$$\sqrt{k_n}(\hat{\theta}_n^{(2)}(\boldsymbol{\tau}_i)-\theta(\boldsymbol{\tau}_i))_{i=1,\ldots,m} \overset{D}{\to} (\Theta(\boldsymbol{\tau}_i \varkappa(-\ln\tilde{H}(\boldsymbol{\tau}_i))^{-1}))_{i=1,\ldots,m}.$$

Although the estimators are very different from the point of view of their construction, they share the same asymptotic distribution up to a proportional factor.

**Example 2.1 (continued).** The calculation of the asymptotic variance of $\hat{\theta}_n^{(1)}(\boldsymbol{\tau})$ and $\hat{\theta}_n^{(2)}(\boldsymbol{\tau})$ can be carried out explicitly. Let $M(\boldsymbol{\tau}) = -\ln\tilde{H}(\boldsymbol{\tau})L^{-1}(\boldsymbol{\tau})$. We have

$$\mathrm{Var}(\Theta(\boldsymbol{\tau} L^{-1}(\boldsymbol{\tau}))) = M(\boldsymbol{\tau})^{-2}(\mathrm{e}^{M(\boldsymbol{\tau})} - 1 - M(\boldsymbol{\tau})),$$

$$\mathrm{Var}(\Theta(\boldsymbol{\tau}\varkappa(-\ln\tilde{H}(\boldsymbol{\tau}))^{-1})) = (\varkappa^{-2}(\mathrm{e}^\varkappa - 1 - \varkappa)).$$

It is worth mentioning that the asymptotic variance of $\hat{\theta}_n^{(2)}(\boldsymbol{\tau})$ does not depend on $\boldsymbol{\tau}$. It is smaller than the asymptotic variance of $\hat{\theta}_n^{(1)}(\boldsymbol{\tau})$ if $\varkappa \leq M(\boldsymbol{\tau})$.

Note that if $\boldsymbol{\tau} = \boldsymbol{\tau}^{(i)}$, $L=1$ and $\varkappa = \tau_i$, then we obtain the same asymptotic variance as for $\hat{\theta}_{1,n}^{(\tau_i)}$ (see Remark 3.1).

**Example 2.2 (continued).** Let us characterize the asymptotic variance of the second estimator. We have

$$\mathrm{Var}(\Theta(\boldsymbol{\tau}\varkappa(-\ln\tilde{H}(\boldsymbol{\tau}))^{-1}))$$
$$= \varkappa^{-2}\left(\exp\left(\varkappa\frac{\theta_1\tau_1+\theta_2\tau_2}{\tau_1+\tau_2}\right) - 2\varkappa\frac{\theta_1\tau_1+\theta_2\tau_2}{\tau_1+\tau_2} - 1\right)$$
$$+ \varkappa^{-1}\frac{\theta_1^3\tau_1\sum_{j=1}^\infty j^2\pi_1(j) + \theta_2^3\tau_2\sum_{j=1}^\infty j^2\pi_2(j)}{\tau_1+\tau_2}$$

$$+ \varkappa^{-1}\theta_1\theta_2 \frac{\theta_1\tau_1 + \theta_2\tau_2}{\tau_1 + \tau_2} \sum_{\substack{i\geq 0, j\geq 0, k\geq 0 \\ i+j+k\geq 1}} (i+k)(j+l)(\pi^{(\boldsymbol{\tau}^{(1)},\boldsymbol{\tau}^{(2)})} + \pi^{(\boldsymbol{\tau}^{(2)},\boldsymbol{\tau}^{(1)})})(i,j,k).$$

For the first estimator, replace $\varkappa$ by $M(\boldsymbol{\tau})$. A comparison between the two asymptotic variances is not obvious.

Note that if $\boldsymbol{\tau} = \boldsymbol{\tau}^{(i)}$, $L = 1$ and $\varkappa = \tau_i$, then we obtain the same asymptotic variance as for $\hat{\theta}_{1,n}^{(\tau_i)}$ (see Remark 3.1).

It is possible to weaken Condition (C3)(ii) by assuming that there exists an open set $O$ included in $(0,1)^d$ where the functions $H$ and $\tilde{H}$ are (Fréchet) differentiable. One then has to replace $\Psi_L$ by $\{\boldsymbol{\tau} : \boldsymbol{\tau}/L(\boldsymbol{\tau}) \in O \cap [0,1]^d \setminus \{\mathbf{0}\}\}$ and $\Psi_\varkappa$ by $\{\boldsymbol{\tau} : \varkappa\boldsymbol{\tau}/(-\ln\tilde{H}(\boldsymbol{\tau})) \in O \cap [0,1]^d \setminus \{\mathbf{0}\}\}$ in Theorem 4.1.

## 5. Simulation study

In this section, a simulation study is conducted to investigate the performance of the estimators on samples of moderate size. Data are simulated from:

- the bivariate independent process with independent components of Example 2.1 – we have $\theta(\boldsymbol{\tau}) = 1$;
- the bivariate squared ARCH(1) process with independent components of Example 2.2 – we choose $\eta = 2 \times 10^{-5}$, $\lambda_1 = 0.7$, $\lambda_2 = 0.3$ so then we have (see [17])

$$\theta(\boldsymbol{\tau}) = \frac{0.579\tau_1 + 0.887\tau_2}{\tau_1 + \tau_2};$$

- the bivariate autoregressive process of order 1 with dependent innovations of Example 2.3. We choose $\rho_1 = \rho_2 = 1/2$ and $B(x_1, x_2) = (x_1^{-2} + x_2^{-2})^{1/2}$ so then we have $\theta(\boldsymbol{\tau}) = 1/2$.

We study the performances of the first estimator $\hat{\theta}_n^{(1)}$ associated with the functions $L_{c,a}(\boldsymbol{\tau}) = c(\sum_{i=1}^2 |\tau_i|^a)^{1/a}$ for $c = 2$ and $a = 1$, $c = 1$ and $a = 1$, $c = 2$ and $a = 2$, $c = 1$ and $a = 2$, and we undertake comparisons with the second estimator $\hat{\theta}_n^{(2)}$ associated with $\varkappa = 1$. For each process, we generate 500 sequences of length $n = 2000$ and for each sequence, we compute the estimates for $\boldsymbol{\tau} = (\cos\phi, \sin\phi)$ with $\phi = k\pi/22$ and $k = 1, \ldots, 10$.

Figures 1, 2 and 3 show the means (left) and the root mean squared errors (RMSE) (right) of the estimates as functions of the angle, $\phi$, for $k_n = 50, 100, 150, 200$ and for the three processes. First, observe that the bias of the estimators decreases as the size of the blocks increases. The estimators are nearly unbiased for $k_n = 50$ and $k_n = 100$, but they show a positive bias when $k_n = 200$, except in the case of the bivariate squared ARCH process and for the large values of $\phi$. Conversely, the variances of the estimators increase as the size of the blocks increases. This is the ordinary variance-bias trade-off encountered

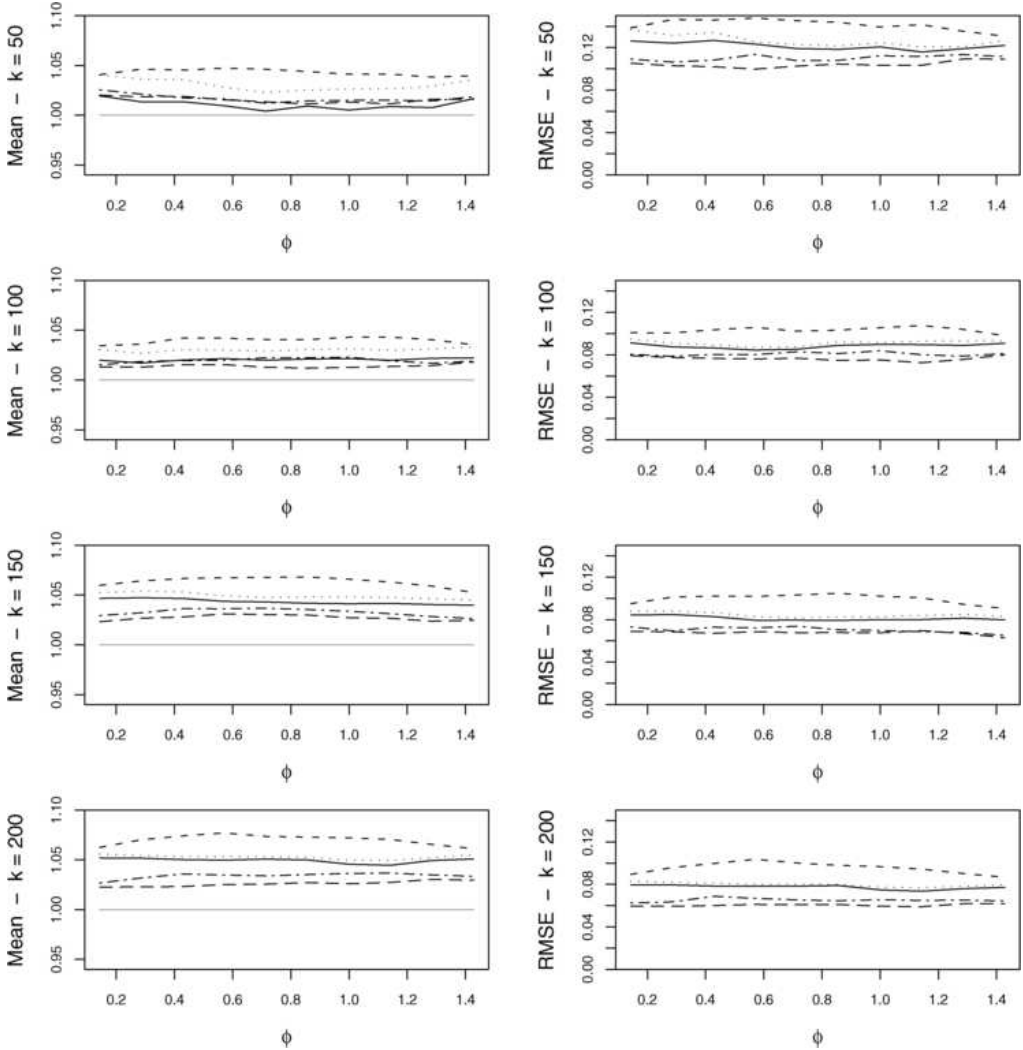

**Figure 1.** The i.i.d. sequence. Left: means of the estimates of the multivariate extremal index function; the gray solid line represents the true function. Right: RMSE of the estimates of the multivariate extremal index function. The estimators which are considered are $\hat{\theta}_n^{(1)}$ associated with $L_{2,1}$ (– – –), $L_{1,1}$ (· · · ·), $L_{2,2}$ (- – - –), $L_{1,2}$ (- - -) and $\hat{\theta}_n^{(2)}$ associated with $\varkappa = 1$ (——). The graphs show the average over 500 samples.

with the blocks estimators. Note that the minimum of the RMSE is generally observed for large values of $k_n$ (150 or 200).

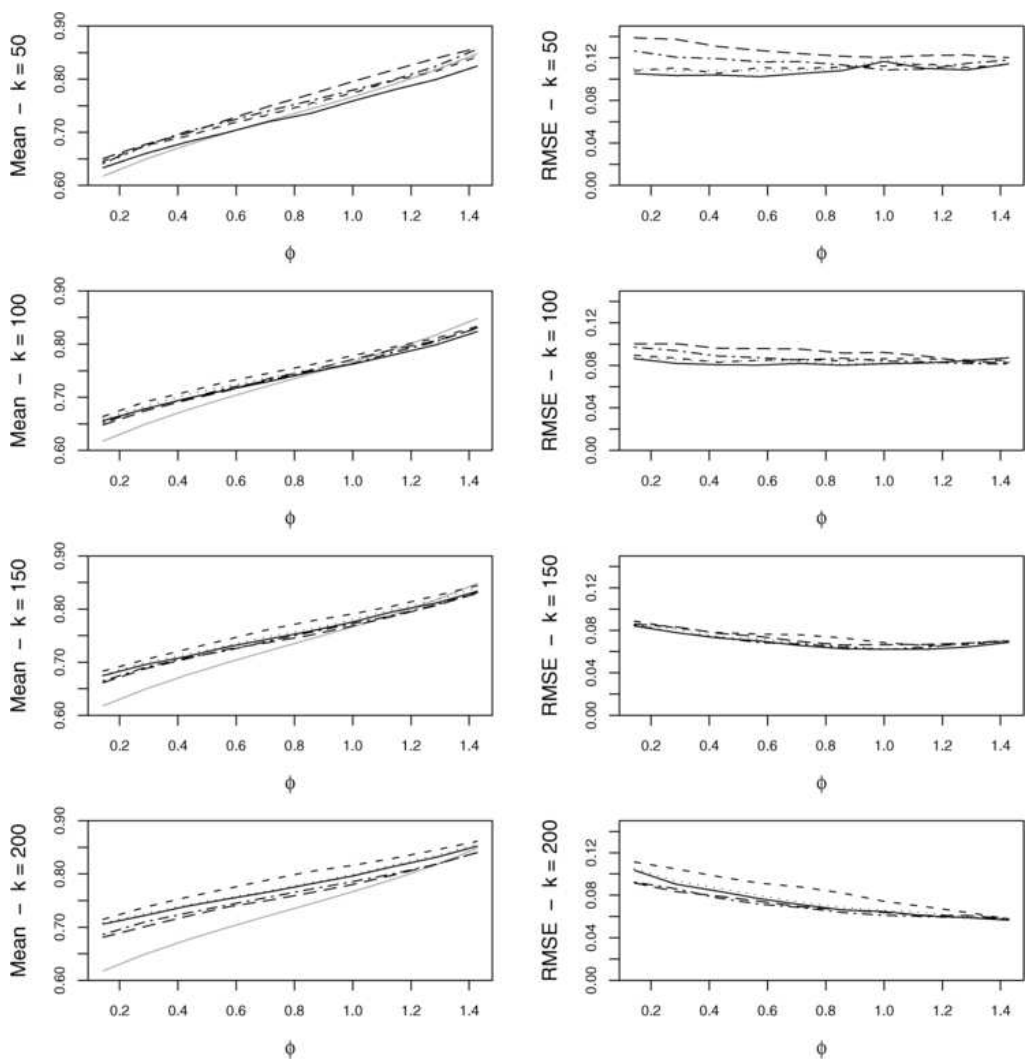

**Figure 2.** The bivariate squared ARCH(1) process. Left: means of the estimates of the multivariate extremal index function; the gray solid line represents the true function; Right: RMSE of the estimates of the multivariate extremal index function; the estimators which are considered are $\hat{\theta}_n^{(1)}$ associated with $L_{2,1}$ (– – –), $L_{1,1}$ (· · · ·), $L_{2,2}$ (- – - –), $L_{1,2}$ (- - -) and $\hat{\theta}_n^{(2)}$ associated with $\varkappa = 1$ (——). The graphs show the average over 500 samples.

For the i.i.d. sequence, the first estimator associated with the function $L_{2,1}$ performs uniformly better. The reasons for this may be that the asymptotic variance is smaller and that the estimated thresholds are higher than those with the choice $c = 1$ and hence the bias is smaller. The RMSE is mimimal for $k_n = 200$.

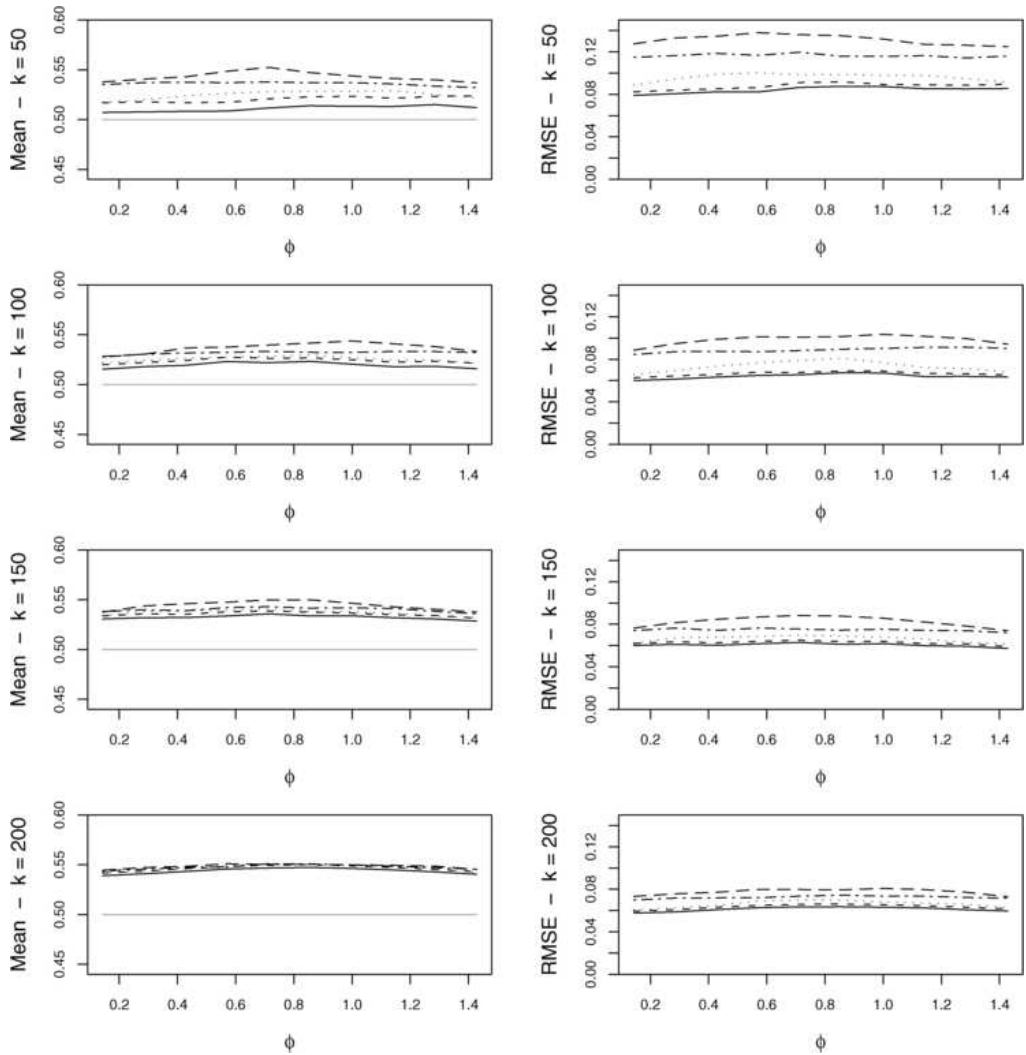

**Figure 3.** The bivariate autoregressive process. Left: means of the estimates of the multivariate extremal index function; the gray solid line represents the true function. Right: RMSE of the estimates of the multivariate extremal index function. The estimators which are considered are $\hat{\theta}_n^{(1)}$ associated with $L_{2,1}$ (– – –), $L_{1,1}$ (· · · ·), $L_{2,2}$ (- – – -), $L_{1,2}$ (- - -) and $\hat{\theta}_n^{(2)}$ associated with $\varkappa = 1$ (——). The graphs show the average over 500 samples.

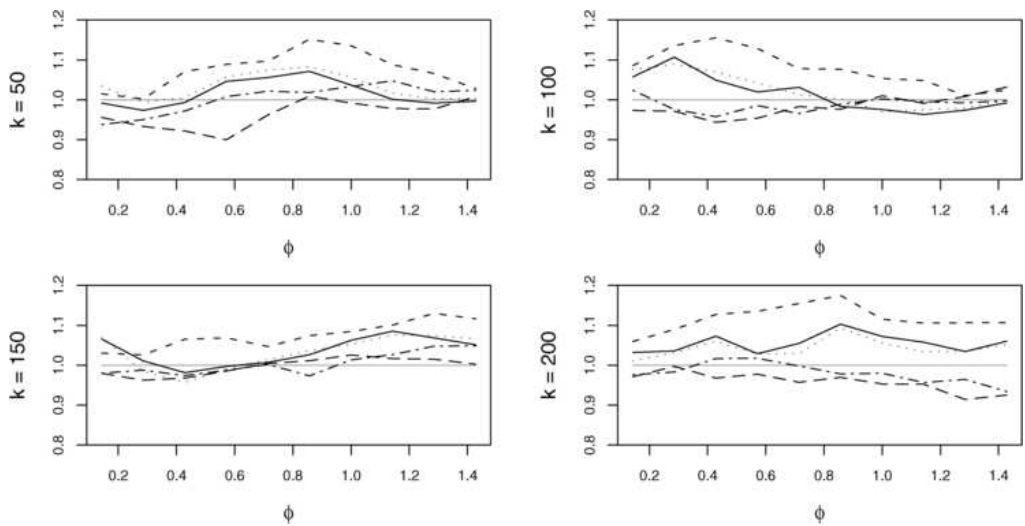

**Figure 4.** The i.i.d. sequence. Ratios between the sample variances and the asymptotic variances for the estimators of the multivariate extremal index function are shown. The estimators which are considered are $\hat{\theta}_n^{(1)}$ associated with $L_{2,1}$ (– – –), $L_{1,1}$ (· · · ·), $L_{2,2}$ (- – – -), $L_{1,2}$ (- - -) and $\hat{\theta}_n^{(2)}$ associated with $\varkappa = 1$ (——). The graphs show the average over 1000 samples.

For the bivariate squared ARCH(1) process, $\hat{\theta}_n^{(1)}$ associated with the function $L_{1,1}$ and $\hat{\theta}_n^{(2)}$ perform better than the other estimators. Note that, as for the previous process, $-\ln \tilde{H}$ and $L_{1,1}$ are equal, which explains why both estimators have the same asymptotic variance. The RMSE is mimimal for $k_n = 150$, except for the large values of $\phi$.

For the autoregressive process, there is a relatively small sensivity of the estimates to the choice of the estimator when $k_n = 200$. The second estimator always performs better.

Overall, $\hat{\theta}_n^{(2)}$ appears as a good candidate to estimate the multivariate extremal index function. Its performance on samples of moderate size is often better than the performance of $\hat{\theta}_n^{(1)}$ and, moreover, it does not necessitate the choice of a tuning function.

Figure 4 shows, for the first process, the ratios between the sample variances and the asymptotic variances for the estimators of the multivariate extremal index function. It illustrates that for sequences of length at least $n = 2000$, the variances of the estimators can be well approximated by the asymptotic variances when they can be calculated or estimated.

## 6. Intermediate results and proof of Theorem 4.1

The proof of Theorem 4.1 and some results related to the weak convergence of intermediate empirical processes are gathered in this section. We let $K$ be a generic constant whose value may change from appearance to appearance.

We first introduce the Skorokhod space of ladcag multiparameter functions and give the essential ingredients that will be required for characterizing the asymptotic behavior of the intermediate processes.

Let $\boldsymbol{B}_d$ be a cube in $\mathbb{R}^d$. If $\boldsymbol{\tau} \in \boldsymbol{B}_d$ and if, for $i=1,\ldots,d$, $R_i$ is one of the relations $\leq$ and $>$, then let $Q_{R_1,\ldots,R_d}(\boldsymbol{\tau})$ be the quadrant

$$\{\boldsymbol{\sigma} = (\sigma_1,\ldots,\sigma_d) \in \boldsymbol{B}_d : \sigma_i R_i \tau_i, i=1,\ldots,d\}.$$

We denote by $D(\boldsymbol{B}_d)$ the space of functions from $\boldsymbol{B}_d$ to $\mathbb{R}$ which are "continuous from below, with limits from above" in the sense defined by [1]. More precisely, $f \in D(\boldsymbol{B}_d)$ if, for each $\boldsymbol{\tau} \in \boldsymbol{B}_d$, $f_Q(\boldsymbol{\tau}) = \lim_{\boldsymbol{\sigma} \to \boldsymbol{\tau}, \boldsymbol{\sigma} \in Q} f(\boldsymbol{\sigma})$ exists for each of the $2^d$ quadrants $Q = Q_{R_1,\ldots,R_d}(\boldsymbol{\tau})$ and $f(\boldsymbol{\tau}) = f_{Q_{\leq,\ldots,\leq}}$. Let us assume that it is equipped with the metric $d^o$ which is equivalent to the Skorohod metric $d$ and such that it makes $D(\boldsymbol{B}_d)$ a complete separable metric space (see [1], Section 2 and [3], Section 12).

A sequence $(f_n)_{n\geq 1}$ of $D(\boldsymbol{B}_d)$-valued processes converges weakly in the Skorohod topology to a $D(\boldsymbol{B}_d)$-valued process $f$ ($f_n(\cdot) \Rightarrow f(\cdot)$) if $\mathbb{E}\varphi(f_n) \to \mathbb{E}\varphi(f_n)$ for all Skorohod-continuous bounded functions $\varphi : D(\boldsymbol{B}_d) \to \mathbb{R}$. A criterion for the weak convergence of $D(\boldsymbol{B}_d)$-valued processes can be given in terms of the weak convergence of the corresponding finite-dimensional distributions together with a tightness condition (see Theorem 1 in [1] and the proof of Theorem 6.1).

It is often convenient to consider the restrictions of the functions of $D(\boldsymbol{B}_d)$ to a subset of $\boldsymbol{B}_d$. If $\boldsymbol{C}_d$ is a cube included in $\boldsymbol{B}_d$ and if $f \in D(\boldsymbol{B}_d)$, we denote by $r_{\boldsymbol{C}_d} f$ the restriction of $f$ to $\boldsymbol{C}_d$. We have the following convergence property: if $f_n(\cdot) \Rightarrow f(\cdot)$ in $D(\boldsymbol{B}_d)$ and $f$ is continuous at the lower boundary of $\boldsymbol{C}_d$, then $r_{\boldsymbol{C}_d} f_n(\cdot) \Rightarrow r_{\boldsymbol{C}_d} f(\cdot)$ in $D(\boldsymbol{C}_d)$ (see, e.g., Lemma 4.17 in [35] for the univariate case).

We now turn to the definition of the intermediate processes and characterize their asymptotic distribution. First, let us introduce, for $\boldsymbol{\tau} \in [0,1]^d$,

$$V_n(\boldsymbol{\tau}) = \sqrt{k_n}(H_n(\boldsymbol{\tau}) - P(N_{r_n,1}^{(\boldsymbol{\tau})} = 0)),$$

$$W_n(\boldsymbol{\tau}) = \sqrt{k_n}((-\ln \tilde{H}_n(\boldsymbol{\tau})) - r_n P(\boldsymbol{X}_l \not\leq \boldsymbol{u}_{r_n}(\boldsymbol{\tau}))),$$

where

$$H_n(\boldsymbol{\tau}) = \frac{1}{k_n} \sum_{j=1}^{k_n} 1_{\{N_{r_n,j}^{(\boldsymbol{\tau})} = 0\}} \quad \text{and} \quad -\ln \tilde{H}_n(\boldsymbol{\tau}) = \frac{1}{k_n} \sum_{j=1}^{k_n} N_{r_n,j}^{(\boldsymbol{\tau})}.$$

We define the first intermediate $D([0,1]^d) \times D([0,1]^d)$-valued process by $\boldsymbol{U}_n(\boldsymbol{\tau}) = (V_n(\boldsymbol{\tau}), W_n(\boldsymbol{\tau}))'$. Observe that $\boldsymbol{U}_n$ depends on the unknown vector of thresholds $\boldsymbol{u}_{r_n}(\boldsymbol{\tau})$ and cannot be used in practice. In the univariate case, $W_n$ is called the *tail empirical process* and has been studied for dependent sequences in [7, 8] and [38].

**Theorem 6.1.** *Suppose that (C1) and (C2) hold. There exists a pathwise continuous centered Gaussian process $\boldsymbol{U}$ with covariance matrix $\boldsymbol{C}(\cdot,\cdot) = (C_{i,j}(\cdot,\cdot))_{1\leq i,j\leq 2}$ given in Appendix B such that $\boldsymbol{U}_n(\cdot) \Rightarrow \boldsymbol{U}(\cdot) \equiv (V(\cdot), W(\cdot))'$ in $D([0,1]^d) \times D([0,1]^d)$.*

The proof of Theorem 6.1 is presented as a series of two lemmas. Let us define the large blocks $I_j^\triangle$ and the small blocks $I_j^*$ by, for $j = 1, \ldots, k_n$,

$$I_j^\triangle = [(j-1)r_n + 1, \ldots, jr_n - l_n], \qquad I_j^* = [jr_n - l_n + 1, \ldots, jr_n].$$

We introduce the quantities

$$N_{r_n,j}^{(\boldsymbol{\tau}),\triangle} = \sum_{l \in I_j^\triangle} 1_{\{\boldsymbol{X}_l \not\leq \boldsymbol{u}_{r_n}(\boldsymbol{\tau})\}}, \qquad N_{r_n,j}^{(\boldsymbol{\tau}),*} = \sum_{l \in I_j^*} 1_{\{\boldsymbol{X}_l \not\leq \boldsymbol{u}_{r_n}(\boldsymbol{\tau})\}}, \qquad j = 1, \ldots, k_n,$$

$$H_n^\triangle(\boldsymbol{\tau}) = \frac{1}{k_n} \sum_{j=1}^{k_n} 1_{\{N_{r_n,j}^{(\boldsymbol{\tau}),\triangle}=0\}}, \qquad H_n^*(\boldsymbol{\tau}) = -\frac{1}{k_n} \sum_{j=1}^{k_n} 1_{\{N_{r_n,j}^{(\boldsymbol{\tau}),\triangle}=0, N_{r_n,j}^{(\boldsymbol{\tau}),*}>0\}},$$

$$-\ln \tilde{H}_n^\triangle(\boldsymbol{\tau}) = \frac{1}{k_n} \sum_{j=1}^{k_n} N_{r_n,j}^{(\boldsymbol{\tau}),\triangle}, \qquad -\ln \tilde{H}_n^*(\boldsymbol{\tau}) = \frac{1}{k_n} \sum_{j=1}^{k_n} N_{r_n,j}^{(\boldsymbol{\tau}),*},$$

and consider the following processes

$$V_n^\Delta(\boldsymbol{\tau}) = \sqrt{k_n}(H_n^\triangle(\boldsymbol{\tau}) - P(N_{r_n,1}^{(\boldsymbol{\tau}),\triangle} = 0));$$
$$V_n^*(\boldsymbol{\tau}) = \sqrt{k_n}(H_n^*(\boldsymbol{\tau}) + P(N_{r_n,j}^{(\boldsymbol{\tau}),\triangle} = 0, N_{r_n,j}^{(\boldsymbol{\tau}),*} > 0));$$
$$W_n^\Delta(\boldsymbol{\tau}) = \sqrt{k_n}((-\ln \tilde{H}_n^\triangle(\boldsymbol{\tau})) - (r_n - l_n)P(\boldsymbol{X}_l \not\leq \boldsymbol{u}_{r_n}(\boldsymbol{\tau})));$$
$$W_n^*(\boldsymbol{\tau}) = \sqrt{k_n}((-\ln \tilde{H}_n^*(\boldsymbol{\tau})) - l_n P(\boldsymbol{X}_l \not\leq \boldsymbol{u}_{r_n}(\boldsymbol{\tau})));$$
$$\boldsymbol{U}_n^\Delta(\boldsymbol{\tau}) = (V_n^\Delta(\boldsymbol{\tau}), W_n^\Delta(\boldsymbol{\tau}))', \qquad \boldsymbol{U}_n^*(\boldsymbol{\tau}) = (V_n^*(\boldsymbol{\tau}), W_n^*(\boldsymbol{\tau}))'.$$

Note that $H_n(\boldsymbol{\tau}) = H_n^\triangle(\boldsymbol{\tau}) + H_n^*(\boldsymbol{\tau})$, $-\ln \tilde{H}_n(\boldsymbol{\tau}) = -\ln \tilde{H}_n^\triangle(\boldsymbol{\tau}) - \ln \tilde{H}_n^*(\boldsymbol{\tau})$, $V_n(\boldsymbol{\tau}) = V_n^\Delta(\boldsymbol{\tau}) + V_n^*(\boldsymbol{\tau})$, $W_n(\boldsymbol{\tau}) = W_n^\Delta(\boldsymbol{\tau}) + W_n^*(\boldsymbol{\tau})$ and $\boldsymbol{U}_n(\boldsymbol{\tau}) = \boldsymbol{U}_n^\Delta(\boldsymbol{\tau}) + \boldsymbol{U}_n^*(\boldsymbol{\tau})$.

**Lemma 6.1.** *Suppose that (C1) and (C2) hold. Let $m \geq 1$ and $\boldsymbol{\tau}_1, \ldots, \boldsymbol{\tau}_m \in [0,1]^d$. Then,*

$$(\boldsymbol{U}_n(\boldsymbol{\tau}_i))_{i=1,\ldots,m} \xrightarrow{D} (\boldsymbol{U}(\boldsymbol{\tau}_i))_{i=1,\ldots,m}.$$

**Proof.** By similar arguments as in Lemma 6.6 of [37] with

$$4d \vee \frac{2\omega}{\omega - 1} < v < r$$

(this can always be assumed), we have $\boldsymbol{U}_n^*(\boldsymbol{\tau}) \xrightarrow{P} 0$. It is only needed to be checked that

$$(\boldsymbol{U}_n^\Delta(\boldsymbol{\tau}_i))_{i=1,\ldots,m} \xrightarrow{D} (\boldsymbol{U}(\boldsymbol{\tau}_i))_{i=1,\ldots,m}.$$

But we can use the arguments of Lemma 6.7 in [37] and replace (C0)(b) in [37] by (C2)(a) with $\boldsymbol{\tau}_2 = 0$ in order to establish the weak convergence and to conclude that $(\boldsymbol{U}(\boldsymbol{\tau}_i))_{i=1,\ldots,m}$ is a Gaussian centered random vector with covariance matrix

$$\mathrm{Cov}(\boldsymbol{U}(\boldsymbol{\tau}_l), \boldsymbol{U}(\boldsymbol{\tau}_k)) = \mathbb{E}(\boldsymbol{U}(\boldsymbol{\tau}_l)\boldsymbol{U}'(\boldsymbol{\tau}_k)) = \boldsymbol{C}(\boldsymbol{\tau}_l, \boldsymbol{\tau}_k), \qquad 1 \leq l, k \leq m. \qquad \square$$

**Lemma 6.2.** *Suppose that (C1) and (C2) hold. Let us define the modulus of continuity of $f \in D([0,1]^d)$ by*

$$w_f(\delta) = \sup\{|f(\boldsymbol{\tau}) - f(\boldsymbol{\tau}')| : \boldsymbol{\tau}, \boldsymbol{\tau}' \in [0,1]^d, |\boldsymbol{\tau} - \boldsymbol{\tau}'| < \delta\}.$$

*Let $\varepsilon > 0$. Then,*

$$\lim_{\delta \to 0} \limsup_n P(w_{V_n}(\delta) > \varepsilon) = 0, \tag{6.1}$$

$$\lim_{\delta \to 0} \limsup_n P(w_{W_n}(\delta) > \varepsilon) = 0. \tag{6.2}$$

**Proof.** We combine some arguments from Section 5 of [29] and some arguments from the proof of Theorem 1 in [5]. Let $L(2^{m_n})$ be the set of all points $(l_1, \ldots, l_d)/2^{m_n}$ with $l_i \in \{0, 1, \ldots, 2^{m_n}\}$, $i = 1, \ldots, d$.

Since $H_n(\boldsymbol{\tau})$ is a monotonically non-increasing function in each component of $\boldsymbol{\tau}$ and $-\ln \tilde{H}_n(\boldsymbol{\tau})$ is a monotonically non-decreasing function in each component of $\boldsymbol{\tau}$, we have (see [31], page 262)

$$w_{V_n}(\delta) \leq 6 \sup |V_n(\boldsymbol{\tau}_1) - V_n(\boldsymbol{\tau}_2)| + 4d\sqrt{k_n} 2^{-m_n},$$
$$w_{W_n}(\delta) \leq 6 \sup |W_n(\boldsymbol{\tau}_1) - W_n(\boldsymbol{\tau}_2)| + 4d\sqrt{k_n} 2^{-m_n},$$

where the "sup" is to be taken over all $\boldsymbol{\tau}_1, \boldsymbol{\tau}_2 \in L(2^{m_n})$ with $|\boldsymbol{\tau}_1 - \boldsymbol{\tau}_2| \leq \delta + 2^{-m_n+1}$. If $m_n$ is chosen such that $\lim_{n \to \infty} \sqrt{k_n} 2^{-m_n} = 0$, (6.1) and (6.2) will follow if we can show that

$$\lim_{\delta \to 0} \limsup_n P(\sup |V_n(\boldsymbol{\tau}_1) - V_n(\boldsymbol{\tau}_2)| > \varepsilon) = 0,$$

$$\lim_{\delta \to 0} \limsup_n P(\sup |W_n(\boldsymbol{\tau}_1) - W_n(\boldsymbol{\tau}_2)| > \varepsilon) = 0,$$

where the "sup" is to be taken over all $\boldsymbol{\tau}_1, \boldsymbol{\tau}_2 \in L(2^{m_n})$ with $|\boldsymbol{\tau}_1 - \boldsymbol{\tau}_2| \leq \delta$. Let us define

$$Y_{V,j}(\boldsymbol{\tau}_1, \boldsymbol{\tau}_2) = (1_{\{N_{r_n,j}^{(\boldsymbol{\tau}_1)} = 0\}} - P(N_{r_n,1}^{(\boldsymbol{\tau}_1)} = 0)) - (1_{\{N_{r_n,j}^{(\boldsymbol{\tau}_2)} = 0\}} - P(N_{r_n,1}^{(\boldsymbol{\tau}_2)} = 0)),$$

$$Y_{W,j}(\boldsymbol{\tau}_1, \boldsymbol{\tau}_2) = (N_{r_n,j}^{(\boldsymbol{\tau}_1)} - r_n P(\boldsymbol{X}_l \not\leq \boldsymbol{u}_{r_n}(\boldsymbol{\tau}_1))) - (N_{r_n,j}^{(\boldsymbol{\tau}_2)} - r_n P(\boldsymbol{X}_l \not\leq \boldsymbol{u}_{r_n}(\boldsymbol{\tau}_2))),$$

$$S_{V,n}(\boldsymbol{\tau}_1,\boldsymbol{\tau}_2) = \sum_{j=1}^{k_n} Y_{V,j}(\boldsymbol{\tau}_1,\boldsymbol{\tau}_2), \qquad S_{W,n}(\boldsymbol{\tau}_1,\boldsymbol{\tau}_2) = \sum_{j=1}^{k_n} Y_{W,j}(\boldsymbol{\tau}_1,\boldsymbol{\tau}_2).$$

We now want to use equation (4.3) of Theorem 4.1 in [39]. Note that $\alpha_{r_n l} \leq \alpha_l$ for $l \geq 1$ and $r_n \geq 1$. Let $2 < v < p < r \leq \infty$, $\kappa > 0$, and assume that $\omega > v/(v-2)$ and $\omega \geq (p-1)r/(r-p)$. We choose $p = 4d$ and $r = 2dv$ with $2 < v < 4$, which leads to $\omega \geq (4d-1)r/(r-4d)$. We deduce that

$$\mathbb{E}|S_{V,n}(\boldsymbol{\tau}_1,\boldsymbol{\tau}_2)|^{4d} \leq k_n^{2d}\|Y_{V,j}(\boldsymbol{\tau}_1,\boldsymbol{\tau}_2)\|_v^{4d} + k_n^{1+\kappa}\|Y_{V,j}(\boldsymbol{\tau}_1,\boldsymbol{\tau}_2)\|_{2dv}^{4d},$$

$$\mathbb{E}|S_{W,n}(\boldsymbol{\tau}_1,\boldsymbol{\tau}_2)|^{4d} \leq k_n^{2d}\|Y_{W,j}(\boldsymbol{\tau}_1,\boldsymbol{\tau}_2)\|_v^{4d} + k_n^{1+\kappa}\|Y_{W,j}(\boldsymbol{\tau}_1,\boldsymbol{\tau}_2)\|_{2dv}^{4d}.$$

Note that, for $\lambda \geq 1$,

$$|Y_{V,j}(\boldsymbol{\tau}_1,\boldsymbol{\tau}_2)|^\lambda \leq 2^\lambda(|1_{\{N_{r_n,j}^{(\boldsymbol{\tau}_1)}=0\}} - 1_{\{N_{r_n,j}^{(\boldsymbol{\tau}_2)}=0\}}|^\lambda + |P(N_{r_n,j}^{(\boldsymbol{\tau}_1)}=0) - P(N_{r_n,j}^{(\boldsymbol{\tau}_2)}=0)|^\lambda),$$

$$|Y_{W,j}(\boldsymbol{\tau}_1,\boldsymbol{\tau}_2)|^\lambda \leq 2^\lambda(|N_{r_n,1}^{(\boldsymbol{\tau}_1)} - N_{r_n,1}^{(\boldsymbol{\tau}_2)}|^\lambda + |\mathbb{E}N_{r_n,1}^{(\boldsymbol{\tau}_1)} - \mathbb{E}N_{r_n,1}^{(\boldsymbol{\tau}_2)}|^\lambda).$$

Since

$$|1_{\{N_{r_n,j}^{(\boldsymbol{\tau}_1)}=0\}} - 1_{\{N_{r_n,j}^{(\boldsymbol{\tau}_2)}=0\}}| \leq |N_{r_n,j}^{(\boldsymbol{\tau}_1)} - N_{r_n,j}^{(\boldsymbol{\tau}_2)}|,$$

we deduce by (C2)(a) that for $1 \leq \lambda \leq r$,

$$\mathbb{E}|1_{\{N_{r_n,j}^{(\boldsymbol{\tau}_1)}=0\}} - 1_{\{N_{r_n,j}^{(\boldsymbol{\tau}_2)}=0\}}|^\lambda \leq \mathbb{E}|N_{r_n,j}^{(\boldsymbol{\tau}_1)} - N_{r_n,j}^{(\boldsymbol{\tau}_2)}|^\lambda \leq D|\boldsymbol{\tau}_1 - \boldsymbol{\tau}_2|.$$

Moreover,

$$|P(N_{r_n,j}^{(\boldsymbol{\tau}_1)}=0) - P(N_{r_n,j}^{(\boldsymbol{\tau}_2)}=0)| \leq \mathbb{E}|1_{\{N_{r_n,j}^{(\boldsymbol{\tau}_1)}=0\}} - 1_{\{N_{r_n,j}^{(\boldsymbol{\tau}_2)}=0\}}| \leq \mathbb{E}|N_{r_n,1}^{(\boldsymbol{\tau}_1)} - N_{r_n,1}^{(\boldsymbol{\tau}_2)}|$$

and

$$|\mathbb{E}N_{r_n,1}^{(\boldsymbol{\tau}_1)} - \mathbb{E}N_{r_n,1}^{(\boldsymbol{\tau}_2)}| \leq \mathbb{E}|N_{r_n,1}^{(\boldsymbol{\tau}_1)} - N_{r_n,1}^{(\boldsymbol{\tau}_2)}| \leq \mathbb{E}|N_{r_n,1}^{(\boldsymbol{\tau}_1)} - N_{r_n,1}^{(\boldsymbol{\tau}_2)}|^r.$$

It follows that for $\lambda \geq 1$ and $|\boldsymbol{\tau}_1 - \boldsymbol{\tau}_2| < 1$,

$$(|\mathbb{E}N_{r_n,1}^{(\boldsymbol{\tau}_1)} - \mathbb{E}N_{r_n,1}^{(\boldsymbol{\tau}_2)}|)^\lambda \leq (\mathbb{E}|N_{r_n,1}^{(\boldsymbol{\tau}_1)} - N_{r_n,1}^{(\boldsymbol{\tau}_2)}|^r)^\lambda \leq (D|\boldsymbol{\tau}_1-\boldsymbol{\tau}_2|)^\lambda \leq K|\boldsymbol{\tau}_1-\boldsymbol{\tau}_2|.$$

We deduce that for $|\boldsymbol{\tau}_1 - \boldsymbol{\tau}_2| < 1$,

$$\|Y_{V,j}(\boldsymbol{\tau}_1,\boldsymbol{\tau}_2)\|_v^{4d} \leq K|\boldsymbol{\tau}_1-\boldsymbol{\tau}_2|^{4d/v}, \qquad \|Y_{V,j}(\boldsymbol{\tau}_1,\boldsymbol{\tau}_2)\|_{2dv}^{4d} \leq K|\boldsymbol{\tau}_1-\boldsymbol{\tau}_2|^{2/v},$$

$$\|Y_{W,j}(\boldsymbol{\tau}_1,\boldsymbol{\tau}_2)\|_v^{4d} \leq K|\boldsymbol{\tau}_1-\boldsymbol{\tau}_2|^{4d/v}, \qquad \|Y_{W,j}(\boldsymbol{\tau}_1,\boldsymbol{\tau}_2)\|_{2dv}^{4d} \leq K|\boldsymbol{\tau}_1-\boldsymbol{\tau}_2|^{2/v}$$

and it follows that for any $\kappa > 0$,

$$\mathbb{E}|S_{V,n}(\boldsymbol{\tau}_1, \boldsymbol{\tau}_2)|^{4d} \leq K(k_n^{2d}|\boldsymbol{\tau}_1 - \boldsymbol{\tau}_2|^{4d/v} + k_n^{1+\kappa}|\boldsymbol{\tau}_1 - \boldsymbol{\tau}_2|^{2/v}),$$

$$\mathbb{E}|S_{W,n}(\boldsymbol{\tau}_1, \boldsymbol{\tau}_2)|^{4d} \leq K(k_n^{2d}|\boldsymbol{\tau}_1 - \boldsymbol{\tau}_2|^{4d/v} + k_n^{1+\kappa}|\boldsymbol{\tau}_1 - \boldsymbol{\tau}_2|^{2/v}).$$

Since right-hand sides of the previous two inequalities are the same, we only consider the case of the process $W_n$. We then deduce that

$$\mathbb{E}|W_n(\boldsymbol{\tau}_1) - W_n(\boldsymbol{\tau}_2)|^{4d} \leq K(|\boldsymbol{\tau}_1 - \boldsymbol{\tau}_2|^{4d/v} + k_n^{1+\kappa-2d}|\boldsymbol{\tau}_1 - \boldsymbol{\tau}_2|^{2/v}).$$

If $k_n^{1+\kappa-2d} \leq |\boldsymbol{\tau}_1 - \boldsymbol{\tau}_2|^{4d/v-2/v}$ or, equivalently, $|\boldsymbol{\tau}_1 - \boldsymbol{\tau}_2| \geq k_n^{v(1+\kappa-2d)/(2(d-1))}$, then

$$\mathbb{E}|W_n(\boldsymbol{\tau}_1) - W_n(\boldsymbol{\tau}_2)|^{4d} \leq K|\boldsymbol{\tau}_1 - \boldsymbol{\tau}_2|^{4d/v}.$$

In particular, if $\boldsymbol{\tau}_2 = \boldsymbol{\tau}_1 + e_i 2^{-\gamma}$, where $e_i = \mathbf{1}^{(i)}$ such that $2^{-\gamma} \geq k_n^{v(1+\kappa-2d)/(2(d-1))}$, we get

$$\mathbb{E}|W_n(\boldsymbol{\tau}_1) - W_n(\boldsymbol{\tau}_2)|^{4d} \leq K(2^{-r})^{4d/v}.$$

Let $m(\delta) = \max\{\gamma \in \mathbb{N}: \delta 2^\gamma \leq 1\}$ and $0 < a < 1$. By using the same arguments as in Section 5 of [29], we have

$$P(\sup\{|W_n(\boldsymbol{\tau}_1) - W_n(\boldsymbol{\tau}_2)|: \boldsymbol{\tau}_1, \boldsymbol{\tau}_2 \in L(2^{m_n}), |\boldsymbol{\tau}_1 - \boldsymbol{\tau}_2| \leq \delta\} > \varepsilon)$$

$$\leq \sum_{i=1}^d \sum_{\gamma=m(\delta)}^{m_n} \sum_{j_1=1}^{2^\gamma} \cdots \sum_{j_i=0}^{2^\gamma-1} \cdots \sum_{j_d=1}^{2^\gamma} P(|W_n(\boldsymbol{j}) - W_n(\boldsymbol{j} + e_i 2^{-\gamma})|$$

$$> (1-a)a^{\gamma-m(\delta)}\varepsilon(4d^2)^{-1})$$

where $\boldsymbol{j} = (j_1, \ldots, j_d) \in L(2^\gamma)$. If $2^{-\gamma} \geq 2^{-m_n} \geq k_n^{v(1+\kappa-2d)/(2(2d-1))}$, then we get, by Chebyshev's inequality,

$$P(|W_n(\boldsymbol{j}) - W_n(\boldsymbol{j} + e_i 2^{-\gamma})| > (1-a)a^{\gamma-m(\delta)}\varepsilon(4d^2)^{-1})$$

$$\leq \frac{\mathbb{E}|W_n(\boldsymbol{\tau}_1) - W_n(\boldsymbol{\tau}_2)|^{4d}}{((1-a)a^{\gamma-m(\delta)}\varepsilon(4d^2)^{-1})^{4d}} \leq \frac{K(2^{-\gamma})^{4d/v}}{((1-a)a^{\gamma-m(\delta)}\varepsilon(4d^2)^{-1})^{4d}}.$$

It follows that

$$P(\sup\{|W_n(\boldsymbol{\tau}_1) - W_n(\boldsymbol{\tau}_2)|: \boldsymbol{\tau}_1, \boldsymbol{\tau}_2 \in L^{(i)}(2^{m_n}), |\boldsymbol{\tau}_1 - \boldsymbol{\tau}_2| \leq \delta\} > \varepsilon)$$

$$\leq \frac{dK}{((1-a)\varepsilon(4d^2)^{-1})^{4d}} \sum_{\gamma=m(\delta)}^{m_n} (2^{-\gamma})^{d(4/v-1)} \frac{1}{a^{\gamma-m(\delta)}}$$

$$= \frac{dK}{((1-a)\varepsilon(4d^2)^{-1})^{4d}(2^{d(4/v-1)})^{m(\delta)}} \sum_{\gamma=m(\delta)}^{m_n} \frac{1}{(2^{d(4/v-1)}a)^{\gamma-m(\delta)}}.$$

Let us choose $a$ such that $2^{d(4/v-1)}a > 1$, which is possible since $v < 4$, and let us choose $m_n$ such that

$$\lim_{n\to\infty} \sqrt{k_n} 2^{-m_n} = 0 \quad \text{and} \quad \lim_{n\to\infty} k_n^{v(1+\kappa-2d)/(2(2d-1))} 2^{m_n} = 0,$$

that is, such that

$$k_n^{-v(1/2-\kappa/(2(2d-1)))} = o(2^{-m_n}) \quad \text{and} \quad 2^{-m_n} = o(k_n^{-1/2}).$$

This is clearly possible since $v > 2$ and $\kappa$ is arbitrarily small. Now let $m_n$ tend to infinity. The infinite series converges since $2^{d(4/v-1)}a > 1$. Finally, let $\delta$ tend to 0 or, equivalently, $m(\delta)$ tend to infinity. The upper bound tends to zero since $2^{d(4/v-1)} > 1$ and the result follows. □

**Proof of Theorem 6.1.** Lets define the diameter of a rectangle as the length of its shortest side. Call a partition of $[0,1]^d$ formed by finitely many hyperplanes parallel to the coordinate axes a $\delta$-*grid* if each element of the partition is a "right-closed, left-open" rectangle of diameter at least $\delta$ and define $w'_{(\cdot)}(\delta) \colon D([0,1]^d) \to \mathbb{R}$ by

$$w'_f(\delta) = \inf_\Delta \max_{G\in\Delta} \sup_{\boldsymbol{\sigma},\boldsymbol{\tau}\in G} |f(\boldsymbol{\tau}) - f(\boldsymbol{\sigma})|,$$

where the infimum extends over all $\delta$-grids $\Delta$ on $[0,1]^d$. Let us define $\Pi_S \colon D([0,1]^d) \to \mathbb{R}^S$ by $\Pi_S(f) = (f(s))_{s\in S}$ for each finite set $S \subset [0,1]^d$. Let $\mathcal{T}$ be the collection of subsets of $[0,1]^d$ of the form $U_1 \times \cdots \times U_d$ where each $U_j$ contains 0 and 1 and has countable complement. According to Theorem 2 in [1], $V_n \Rightarrow V$ (resp., $W_n \Rightarrow W$) if and only if

 (i) $\Pi_S(V_n) \Rightarrow \Pi_S(V)$ for all finite subsets $S$ of some member of $\mathcal{T}$ (resp., $\Pi_S(W_n) \Rightarrow \Pi_S(W)$);
 (ii) $\lim_{\delta\to 0} \limsup_n P(w'_{V_n}(\delta) > \varepsilon) = 0$ for all $\varepsilon > 0$ (resp., $\lim_{\delta\to 0} \limsup_n P(w'_{V_n}(\delta) > \varepsilon) = 0$).

By Lemma 6.1, we derive the first condition. Now, according to equation (1.7) in [29], we have

$$w'_{V_n}(\delta) \leq w_{V_n}(2\delta) \quad \text{and} \quad w'_{W_n}(\delta) \leq w_{W_n}(2\delta), \qquad 0 < \delta < 1/2.$$

By Lemma 6.2, we derive the second condition. Moreover, by using the same arguments as in the proof of Theorem 15.5 in [2], we can show that $V$ and $W$ belong to $C([0,1]^d)$, the subset of $D([0,1]^d)$ consisting of continuous functions. □

We now substitute the unknown vector of thresholds in the first intermediate process by its estimate, $\hat{\boldsymbol{u}}_{r_n}(\boldsymbol{\tau})$, and replace $P(N^{(\boldsymbol{\tau})}_{r_n,1} = 0)$ and $P(\boldsymbol{X}_l \not\leq \boldsymbol{u}_{r_n}(\boldsymbol{\tau}))$ by their respective limits. Let us introduce

$$\hat{V}_n(\boldsymbol{\tau}) = \sqrt{k_n}(\hat{H}_n(\boldsymbol{\tau}) - H(\boldsymbol{\tau})),$$
$$\hat{W}_n(\boldsymbol{\tau}) = \sqrt{k_n}((-\ln\widehat{\tilde{H}_n}(\boldsymbol{\tau})) - (-\ln\tilde{H}(\boldsymbol{\tau})))$$

and define the second intermediate $D([0,1]^d) \times D([0,1]^d)$-valued process by $\hat{\boldsymbol{U}}_n(\boldsymbol{\tau}) = (\hat{V}_n(\boldsymbol{\tau}), \hat{W}_n(\boldsymbol{\tau}))'$. We now establish the weak convergence of this process.

**Proposition 6.1.** *Suppose that (C1), (C2) and (C3) hold. Then,*

$$\hat{\boldsymbol{U}}_n(\cdot) \Rightarrow \hat{\boldsymbol{U}}(\cdot) \equiv (\hat{V}(\cdot), \hat{W}(\cdot))'$$

*in $D([0,1]^d) \times D([0,1]^d)$, where*

$$\hat{\boldsymbol{U}}(\boldsymbol{\tau}) = \boldsymbol{U}(\boldsymbol{\tau}) + \begin{pmatrix} -\nabla H(\boldsymbol{\tau})' \boldsymbol{Z}(\boldsymbol{\tau}) \\ \tilde{H}^{-1}(\boldsymbol{\tau}) \nabla \tilde{H}(\boldsymbol{\tau})' \boldsymbol{Z}(\boldsymbol{\tau}) \end{pmatrix},$$

$\nabla H(\boldsymbol{\tau}) = (\partial H(\boldsymbol{\tau})/\partial \tau_i)_{i=1,\ldots,d}$, $\nabla \tilde{H}(\boldsymbol{\tau}) = (\partial \tilde{H}(\boldsymbol{\tau})/\partial \tau_i)_{i=1,\ldots,d}$ *and* $\boldsymbol{Z}(\boldsymbol{\tau}) = (W(\pi_i(\boldsymbol{\tau})))_{i=1,\ldots,d}$ *with* $\pi_i(\boldsymbol{\tau}) = \boldsymbol{\tau}^{(i)}$, $i = 1, \ldots, d$.

Note that $\hat{\boldsymbol{U}}$ is well defined on $[0,1]^d \setminus \{\boldsymbol{0}\}$ and can be extended by continuity at $\{\boldsymbol{0}\}$ by setting $\hat{\boldsymbol{U}}(\boldsymbol{0}) = \boldsymbol{U}(\boldsymbol{0}) = (0,0)'$. Moreover, if $\tilde{G}$ has independent components, then $\hat{W} = 0$.

**Proof of Proposition 6.1.** Let us define the functions $\bar{p}_{n,i}$ by

$$\bar{p}_{n,i}(\tau_i) = -\ln(\tilde{H}_n(\boldsymbol{\tau}^{(i)})) = -\ln(\tilde{H}_n((0, \ldots, 0, \tau_i, 0, \ldots, 0))), \qquad i = 1, \ldots, d.$$

The generalized inverse of $\bar{p}_{n,i}$ is given for $0 < \bar{\tau} \leq r_n$ by

$$\bar{p}_{n,i}^{\leftarrow}(\bar{\tau}) = \inf\left\{ \tau \geq 0 : \sum_{j=1}^{r_n k_n} 1_{\{X_{j,i} > F_i^{\leftarrow}(1-\tau/r_n)\}} \geq k_n \bar{\tau} \right\} = r_n \bar{F}_i(X_{(\lceil k_n \bar{\tau} \rceil), i})$$

since $F_i^{\leftarrow}(F_i(X_{(\lceil k_n \bar{\tau} \rceil), i})) = X_{(\lceil k_n \bar{\tau} \rceil), i}$. Without loss of generality, assume that $\bar{p}_{n,i}^{\leftarrow}(0) = 0$. Note that $\bar{p}_{n,i}^{\leftarrow}(\cdot)$ is a caglad function on $[0,1]$. Letting $\tilde{\pi}_i(\boldsymbol{\tau}) = \tau_i$, we have

$$\hat{H}_n(\boldsymbol{\tau}) = H_n(\bar{p}_{n,1}^{\leftarrow}(\tilde{\pi}_1(\boldsymbol{\tau})), \ldots, \bar{p}_{n,d}^{\leftarrow}(\tilde{\pi}_d(\boldsymbol{\tau}))).$$

Let us introduce the functions $\bar{p}_n$, $\bar{p}_n^{\text{inv}}$ and $e_d$ from $[0,1]^d$ to $\mathbb{R}^d$ defined by

$$\bar{p}_n(\boldsymbol{\tau}) = (\bar{p}_{n,1}(\tilde{\pi}_1(\boldsymbol{\tau})), \ldots, \bar{p}_{n,d}(\tilde{\pi}_d(\boldsymbol{\tau})))',$$
$$\bar{p}_n^{\text{inv}}(\boldsymbol{\tau}) = (\bar{p}_{n,1}^{\leftarrow}(\tilde{\pi}_1(\boldsymbol{\tau})), \ldots, \bar{p}_{n,d}^{\leftarrow}(\tilde{\pi}_d(\boldsymbol{\tau})))',$$
$$e_d(\boldsymbol{\tau}) = \boldsymbol{\tau}.$$

By Theorem 6.1, we have $\bar{p}_n(\cdot) \Rightarrow e_d(\cdot)$ in $D([0,1]^d) \times \cdots \times D([0,1]^d)$. It is easily deduced that $\bar{p}_n^{\text{inv}}(\cdot) \Rightarrow e_d(\cdot)$ in $D([0,1]^d) \times \cdots \times D([0,1]^d)$.

Let us define the processes

$$\tilde{V}_n(\boldsymbol{\tau}) = \sqrt{k_n}(H_n(\boldsymbol{\tau}) - H(\boldsymbol{\tau}))$$

$$= V_n(\boldsymbol{\tau}) + \sqrt{k_n}(P(N^{(\boldsymbol{\tau})}_{r_n,1} = 0) - P(N^{(\boldsymbol{\tau})} = 0)),$$

$$\tilde{W}_n(\boldsymbol{\tau}) = \sqrt{k_n}((-\ln \tilde{H}_n(\boldsymbol{\tau})) - (-\ln \tilde{H}(\boldsymbol{\tau})))$$

$$= W_n(\boldsymbol{\tau}) + \sqrt{k_n}(r_n(1 - F(\boldsymbol{u}_{r_n}(\boldsymbol{\tau}))) + \ln \tilde{H}(\boldsymbol{\tau}))$$

and let $\tilde{\boldsymbol{U}}_n(\boldsymbol{\tau}) = (\tilde{V}_n(\boldsymbol{\tau}), \tilde{W}_n(\boldsymbol{\tau}))'$. By (C3)(i), we have

$$\sup_{\boldsymbol{\tau} \in [0,1]^d} |\sqrt{k_n}(P(N^{(\boldsymbol{\tau})}_{r_n,1} = 0) - P(N^{(\boldsymbol{\tau})} = 0))| \to 0 \qquad \text{as } n \to \infty,$$

$$\sup_{\boldsymbol{\tau} \in [0,1]^d} |\sqrt{k_n}(r_n(1 - F(\boldsymbol{u}_{r_n}(\boldsymbol{\tau}))) + \ln \tilde{H}(\boldsymbol{\tau}))| \to 0 \qquad \text{as } n \to \infty$$

and it follows that

$$\tilde{\boldsymbol{U}}_n(\cdot) \Rightarrow \boldsymbol{U}(\cdot)$$

in $D([0,1]^d) \times D([0,1]^d)$. By using the continuous mapping theorem (CMT) and similar arguments as in the beginning of the proof of Theorem 4.2 in [37], we deduce that

$$\tilde{\boldsymbol{U}}_n(\bar{p}_n^{\text{inv}}(\cdot)) \Rightarrow \boldsymbol{U}(\cdot)$$

in $D([0,1]^d) \times D([0,1]^d)$.

Next, note that

$$\hat{V}_n(\boldsymbol{\tau}) = \tilde{V}_n(\bar{p}_n^{\text{inv}}(\boldsymbol{\tau})) + \sqrt{k_n}(H(\bar{p}_n^{\text{inv}}(\boldsymbol{\tau})) - H(\boldsymbol{\tau})),$$

$$\hat{W}_n(\boldsymbol{\tau}) = \tilde{W}_n(\bar{p}_n^{\text{inv}}(\boldsymbol{\tau})) + \sqrt{k_n}(-\ln \tilde{H}(\bar{p}_n^{\text{inv}}(\boldsymbol{\tau})) - (-\ln \tilde{H}(\boldsymbol{\tau}))).$$

Since $\tilde{W}_n \Rightarrow W$ in $D([0,1]^d)$, we have

$$\sqrt{k_n}(\bar{p}_n(\cdot) - e_d(\cdot)) \Rightarrow \boldsymbol{Z}(\cdot)$$

in $D([0,1]^d) \times \cdots \times D([0,1]^d)$. By using Vervaat's lemma [44], we get

$$\sqrt{k_n}(\bar{p}_n^{\text{inv}}(\cdot) - e_d(\cdot)) \Rightarrow -\boldsymbol{Z}(\cdot)$$

in $D([0,1]^d) \times \cdots \times D([0,1]^d)$. We deduce from the differentiability of $H$ and $\tilde{H}$, and the finite increments formula, that

$$\sqrt{k_n} \begin{pmatrix} H(\bar{p}_n^{\text{inv}}(\cdot)) - H(\cdot) \\ -\ln \tilde{H}(\bar{p}_n^{\text{inv}}(\cdot)) - (-\ln \tilde{H}(\cdot)) \end{pmatrix} \Rightarrow \begin{pmatrix} -\nabla H(\cdot)' \boldsymbol{Z}(\cdot) \\ \tilde{H}^{-1}(\cdot) \nabla \tilde{H}(\cdot)' \boldsymbol{Z}(\cdot) \end{pmatrix}$$

in $D([0,1]^d) \times D([0,1]^d)$.

Finally, we get

$$\hat{\boldsymbol{U}}_n(\cdot) \Rightarrow \boldsymbol{U}(\cdot) + \begin{pmatrix} -\nabla H(\cdot)' \boldsymbol{Z}(\cdot) \\ \tilde{H}^{-1}(\cdot) \nabla \tilde{H}(\cdot)' \boldsymbol{Z}(\cdot) \end{pmatrix}$$

in $D([0,1]^d) \times D([0,1]^d)$. □

Let $\bar{\varkappa} = 1/\sup\{(-\ln \tilde{H}(\boldsymbol{\tau}))^{-1} \bigvee_{i=1}^d \tau_i : \boldsymbol{\tau} \in [0,1]^d \setminus \{\mathbf{0}\}\}$ and introduce, for $(\varkappa, \boldsymbol{\tau}) \in [0, \bar{\varkappa}] \times [0,1]^d$,

$$V_n^{\check{Z}}(\varkappa, \boldsymbol{\tau}) = \sqrt{k_n}(H_n^{\check{Z}}(\varkappa, \boldsymbol{\tau}) - e^{-\theta(\boldsymbol{\tau})\varkappa}),$$
$$W_n^{\check{Z}}(\varkappa, \boldsymbol{\tau}) = \sqrt{k_n}(Q_n^{\check{Z}}(\varkappa, \boldsymbol{\tau}) - \varkappa),$$

where

$$H_n^{\check{Z}}(\varkappa, \boldsymbol{\tau}) = \frac{1}{k_n} \sum_{j=1}^{k_n} 1_{\{N_{r_n,j}^{(\varkappa,\boldsymbol{\tau})}=0\}} \quad \text{and} \quad Q_n^{\check{Z}}(\varkappa, \boldsymbol{\tau}) = \frac{1}{k_n} \sum_{j=1}^{k_n} N_{r_n,j}^{(\varkappa,\boldsymbol{\tau})}.$$

We define an additional intermediate $D([0,\bar{\varkappa}]) \times D([0,\bar{\varkappa}])$-valued process by $\boldsymbol{U}_n^{\check{Z}}(\varkappa, \boldsymbol{\tau}) = (V_n^{\check{Z}}(\varkappa, \boldsymbol{\tau}), W_n^{\check{Z}}(\varkappa, \boldsymbol{\tau}))'$, $\varkappa \in [0, \bar{\varkappa}]$. Observe that $\boldsymbol{U}_n^{\check{Z}}$ depends on the estimated series $(\check{Z}_l^{(\boldsymbol{\tau})})_{l \geq 1}$ and on the unknown threshold $v_n^{(\boldsymbol{\tau})}(\varkappa)$. It is worth mentioning that $\boldsymbol{U}_n^{\check{Z}}$ and $\hat{\boldsymbol{U}}_n$ are closely related since

$$\boldsymbol{U}_n^{\check{Z}}(\varkappa, \boldsymbol{\tau}) = \hat{\boldsymbol{U}}_n(\varkappa(-\ln \tilde{H}(\boldsymbol{\tau}))^{-1}\boldsymbol{\tau}).$$

**Corollary 6.1.** *Suppose that (C1), (C2) and (C3) hold. Then for $\boldsymbol{\tau} \in [0,1]^d \setminus \{\mathbf{0}\}$,*

$$\boldsymbol{U}_n^{\check{Z}}((\cdot), \boldsymbol{\tau}) \Rightarrow \hat{\boldsymbol{U}}(\boldsymbol{\tau}(-\ln \tilde{H}(\boldsymbol{\tau}))^{-1}(\cdot))$$

*in $D([0,\bar{\varkappa}]) \times D([0,\bar{\varkappa}])$.*

**Proof.** We have

$$N_{r_n,j}^{(\varkappa,\boldsymbol{\tau})} = \sum_{l \in I_j} 1_{\{\check{Z}_l^{(\boldsymbol{\tau})} > v_{r_n}^{(\boldsymbol{\tau})}(\varkappa)\}} = \sum_{l \in I_j} 1_{\{\max_{i=1,\ldots,d} \tau_i \frac{k_n r_n}{(k_n r_n + 1 - R_{l,i})} > (-\ln \tilde{H}(\boldsymbol{\tau})) r_n / \varkappa\}}$$

$$= \sum_{l \in I_j} 1_{\{\bigcup_{i=1,\ldots,d} (R_{l,i} > k_n r_n + 1 - \tau_i \varkappa(-\ln \tilde{H}(\boldsymbol{\tau}))^{-1} k_n)\}}$$

$$= \sum_{l \in I_j} 1_{\{\bigcup_{i=1,\ldots,d} (X_{l,i} > X_{(\lceil \tau_i \varkappa(-\ln \tilde{H}(\boldsymbol{\tau}))^{-1} k_n \rceil), i)\}} = \hat{N}_{r_n,j}^{(\varkappa(-\ln \tilde{H}(\boldsymbol{\tau}))^{-1}\boldsymbol{\tau})}$$

and it follows that

$$\boldsymbol{U}_n^{\check{Z}}(\varkappa, \boldsymbol{\tau}) = \hat{\boldsymbol{U}}_n(\boldsymbol{\tau}(-\ln \tilde{H}(\boldsymbol{\tau}))^{-1}\varkappa).$$

Fix $\boldsymbol{\tau} \in [0,1]^d \setminus \{\mathbf{0}\}$ and consider the function $\varkappa \mapsto \boldsymbol{U}_n^{\check{Z},\varkappa}(\varkappa, \boldsymbol{\tau})$ from $[0,\bar{\varkappa}]$ to $\mathbb{R}^2$ as an element of $D([0,\bar{\varkappa}]) \times D([0,\bar{\varkappa}])$. Since the map from $D([0,1]^d)$ to $D([0,\bar{\varkappa}])$ taking $f(\cdot)$ to $f((-\ln \tilde{H}(\boldsymbol{\tau}))^{-1}\boldsymbol{\tau}(\cdot))$ is continuous for any $\boldsymbol{\tau} \in [0,1]^d\setminus\{\boldsymbol{0}\}$, we deduce, by the CMT, that

$$\boldsymbol{U}_n^{\check{Z}}((\cdot),\boldsymbol{\tau}) \Rightarrow \hat{\boldsymbol{U}}(\boldsymbol{\tau}(-\ln \tilde{H}(\boldsymbol{\tau}))^{-1}(\cdot))$$

in $D([0,\bar{\varkappa}]) \times D([0,\bar{\varkappa}])$. □

We now derive, from Proposition 6.1 and Corollary 6.1, the distributional asymptotics of the estimators. Let us define, for $\boldsymbol{\tau} \in [0,1]^d\setminus\{\boldsymbol{0}\}$,

$$\tilde{\theta}_n(\boldsymbol{\tau}) = \frac{-\ln \hat{H}_n(\boldsymbol{\tau})}{-\ln \widehat{\tilde{H}_n}(\boldsymbol{\tau})}, \qquad \Theta(\boldsymbol{\tau}) = \frac{1}{\ln \tilde{H}(\boldsymbol{\tau})}(\hat{V}(\boldsymbol{\tau})H^{-1}(\boldsymbol{\tau}) + \hat{W}(\boldsymbol{\tau})\theta(\boldsymbol{\tau})).$$

Note that $\Theta(\cdot)$ has continuous sample paths on $[0,1]^d\setminus\{\boldsymbol{0}\}$.

By *proper cube* we mean a cube included in $[0,1]^d$ which does not contain $\{\boldsymbol{0}\}$.

**Corollary 6.2.** *Suppose that (C1), (C2) and (C3) hold. Let $\boldsymbol{C}_d$ be a proper cube. Then,*

$$\sqrt{k_n}(r_{\boldsymbol{C}_d}\tilde{\theta}_n(\cdot) - r_{\boldsymbol{C}_d}\theta(\cdot)) \Rightarrow r_{\boldsymbol{C}_d}\Theta(\cdot)$$

*in $D(\boldsymbol{C}_d)$.*

**Proof.** We first recall that a map $\Phi$ between topological vector spaces $B_i$, $i=1,2$, is called *Hadamard differentiable tangentially to* some subset $S \subset B_1$ at $f \in B_1$ if there exists a continuous linear map $\nabla\Phi(f)$ from $B_1$ to $B_2$ such that

$$\frac{\Phi(f+t_n g_n) - \Phi(f)}{t_n} \to \nabla\Phi(f) \cdot g$$

for all sequences $t_n \downarrow 0$ and $g_n \in B_1$ converging to $g \in S$. Let $D(\boldsymbol{C}_d, E)$ (resp., $C(\boldsymbol{C}_d, E)$) be the space of functions from $\boldsymbol{C}_d$ to the set $E \subset \mathbb{R}$ which are "continuous from below, with limits from above" (resp., continuous). Let us consider the map $\Phi$ from $D(\boldsymbol{C}_d, (0,1)) \times D(\boldsymbol{C}_d, (0,\infty))$ to $D(\boldsymbol{C}_d, (0,\infty))$ defined by

$$\Phi(f_1, f_2) = \frac{-\ln f_1}{f_2}.$$

Note that this map is Hadamard differentiable tangentially to $C(\boldsymbol{C}_d, \mathbb{R}) \times C(\boldsymbol{C}_d, \mathbb{R})$ at any $(f_1, f_2) \in C(\boldsymbol{C}_d, (0,1)) \times C(\boldsymbol{C}_d, (0,\infty))$. Moreover, $\nabla\Phi(f_1, f_2)$ is defined and continuous on $C(\boldsymbol{C}_d, \mathbb{R}) \times C(\boldsymbol{C}_d, \mathbb{R})$ and is given by

$$\nabla\Phi(f_1, f_2) \cdot (g_1, g_2) = -\frac{1}{f_1 f_2} g_1 + \frac{\ln f_1}{(f_2)^2} g_2.$$

Since

$$r_{\boldsymbol{C}_d}\tilde{\theta}_n(\cdot) = \Phi(r_{\boldsymbol{C}_d}\hat{H}_n(\cdot), r_{\boldsymbol{C}_d}(-\ln \widehat{\tilde{H}_n})(\cdot)),$$

we deduce by the $\delta$-method (see Theorem 3.9.4 in [43]) and Proposition 6.1 that

$$\sqrt{k_n}(r_{\boldsymbol{C}_d}\tilde{\theta}_n(\cdot) - r_{\boldsymbol{C}_d}\theta(\cdot)) \Rightarrow r_{\boldsymbol{C}_d}\Theta(\cdot)$$

in $D(\boldsymbol{C}_d)$. □

We end this section with the proof of Theorem 4.1.

**Proof of Theorem 4.1.** Let $m \geq 1$ and $\boldsymbol{\tau}_1, \ldots, \boldsymbol{\tau}_m \in \Psi_L$. There exists $\boldsymbol{C}_d \subset [0,1]^d$ such that $\boldsymbol{\tau}_1/L(\boldsymbol{\tau}_1), \ldots, \boldsymbol{\tau}_m/L(\boldsymbol{\tau}_m) \in \boldsymbol{C}_d$. By Corollary 6.2, we deduce that

$$\sqrt{k_n}(\hat{\theta}_n^{(1)}(\boldsymbol{\tau}_i) - \theta(\boldsymbol{\tau}_i))_{i=1,\ldots,m} \Rightarrow \left(\Theta\left(\frac{\boldsymbol{\tau}_i}{L(\boldsymbol{\tau}_i)}\right)\right)_{i=1,\ldots,m}.$$

Let $m \geq 1$ and $\boldsymbol{\tau}_1, \ldots, \boldsymbol{\tau}_m \in \Psi_{\varkappa}$. By using similar arguments as for the proof of Corollary 6.1, we have

$$(\boldsymbol{U}_n^{\check{Z}}(\cdot, \boldsymbol{\tau}_i))_{i=1,\ldots,m} \Rightarrow \left(\hat{\boldsymbol{U}}\left(\frac{\boldsymbol{\tau}_i}{(-\ln \tilde{H}(\boldsymbol{\tau}_i))}(\cdot)\right)\right)_{i=1,\ldots,m}$$

in $(D([0,\bar{\varkappa}]))^{2m}$. Let us consider the thresholds

$$\hat{v}_{r_n}^{(\boldsymbol{\tau}_i)}(\varkappa) = \check{Z}_{(\lceil k_n \varkappa \rceil)}^{(\boldsymbol{\tau}_i)} = v_{r_n}^{(\boldsymbol{\tau}_i)}\left(\frac{r_n(-\ln \tilde{H}(\boldsymbol{\tau}_i))}{\check{Z}_{(\lceil k_n \varkappa \rceil)}^{(\boldsymbol{\tau}_i)}}\right), \qquad i = 1, \ldots, m.$$

Recall that

$$\sqrt{k_n}(Q_n^{\check{Z}}((\cdot), \boldsymbol{\tau}_i) - (\cdot)) \Rightarrow \hat{W}\left((\cdot)\frac{\boldsymbol{\tau}_i}{(-\ln \tilde{H}(\boldsymbol{\tau}_i))}\right)$$

in $D([0,\bar{\varkappa}])$. By using Vervaat's lemma [44], we deduce that

$$\sqrt{k_n}\left(\frac{r_n(-\ln \tilde{H}(\boldsymbol{\tau}_i))}{\check{Z}_{(\lceil k_n(\cdot)\rceil)}^{(\boldsymbol{\tau}_i)}} - (\cdot)\right) \Rightarrow -\hat{W}\left((\cdot)\frac{\boldsymbol{\tau}_i}{(-\ln \tilde{H}(\boldsymbol{\tau}_i))}\right)$$

in $D([0,\bar{\varkappa}])$. Note that

$$\sqrt{k_n}\left(\frac{1}{k_n}\sum_{j=1}^{k_n} 1_{\{\hat{N}_{r_n,j}^{(\varkappa,\boldsymbol{\tau})}=0\}} - e^{-\theta(\boldsymbol{\tau}_i)(\cdot)}\right)$$

$$= V_n^{\check{Z}}\left(\frac{r_n(-\ln \tilde{H}(\boldsymbol{\tau}_i))}{\check{Z}_{(\lceil k_n(\cdot)\rceil)}^{(\boldsymbol{\tau}_i)}}, \boldsymbol{\tau}_i\right) + \sqrt{k_n}(e^{-\theta(\boldsymbol{\tau}_i)r_n(-\ln \tilde{H}(\boldsymbol{\tau}_i))/\check{Z}_{(\lceil k_n(\cdot)\rceil)}^{(\boldsymbol{\tau}_i)}} - e^{-\theta(\boldsymbol{\tau}_i)(\cdot)}).$$

We then deduce by the CMT and Corollary 6.1 that

$$V_n^{\check{Z}}\left(\frac{r_n(-\ln\tilde{H}(\boldsymbol{\tau}_i))}{\check{Z}_{(\lceil k_n(\cdot)\rceil)}^{(\boldsymbol{\tau}_i)}},\boldsymbol{\tau}_i\right) \Rightarrow \hat{V}\left((\cdot)\frac{\boldsymbol{\tau}_i}{(-\ln\tilde{H}(\boldsymbol{\tau}_i))}\right)$$

in $D([0,\bar{\varkappa}])$, by the finite increments formula and the CMT that

$$\sqrt{k_n}(e^{-\theta(\boldsymbol{\tau}_i)r_n(-\ln\tilde{H}(\boldsymbol{\tau}_i))/\check{Z}_{(\lceil k_n(\cdot)\rceil)}^{(\boldsymbol{\tau}_i)}} - e^{-\theta(\boldsymbol{\tau}_i)(\cdot)}) \Rightarrow \theta(\boldsymbol{\tau}_i)e^{-\theta(\boldsymbol{\tau}_i)(\cdot)}\hat{W}\left((\cdot)\frac{\boldsymbol{\tau}_i}{(-\ln\tilde{H}(\boldsymbol{\tau}_i))}\right)$$

and by the $\delta$-method that

$$\sqrt{k_n}(\hat{\theta}_n^{(2)}(\boldsymbol{\tau}_i) - \theta(\boldsymbol{\tau}_i))$$
$$\Rightarrow \frac{e^{\theta(\boldsymbol{\tau}_i)(\cdot)}}{-(\cdot)}\left(\hat{V}\left((\cdot)\frac{\boldsymbol{\tau}_i}{(-\ln\tilde{H}(\boldsymbol{\tau}_i))}\right) + \theta(\boldsymbol{\tau}_i)e^{-\theta(\boldsymbol{\tau}_i)(\cdot)}\hat{W}\left((\cdot)\frac{\boldsymbol{\tau}_i}{(-\ln\tilde{H}(\boldsymbol{\tau}_i))}\right)\right)$$
$$= \frac{1}{-(\cdot)}\left(e^{\theta(\boldsymbol{\tau}_i)(\cdot)}\hat{V}\left((\cdot)\frac{\boldsymbol{\tau}_i}{(-\ln\tilde{H}(\boldsymbol{\tau}_i))}\right) + \theta(\boldsymbol{\tau}_i)\hat{W}\left((\cdot)\frac{\boldsymbol{\tau}_i}{(-\ln\tilde{H}(\boldsymbol{\tau}_i))}\right)\right).$$

Since

$$\ln\tilde{H}\left(\varkappa\frac{\boldsymbol{\tau}_i}{(-\ln\tilde{H}(\boldsymbol{\tau}_i))}\right) = -\varkappa \quad \text{and} \quad H\left(\varkappa\frac{\boldsymbol{\tau}_i}{(-\ln\tilde{H}(\boldsymbol{\tau}_i))}\right) = e^{-\theta(\boldsymbol{\tau}_i)\varkappa},$$

we have

$$\frac{1}{-\varkappa}\left(e^{\theta(\boldsymbol{\tau}_i)\varkappa}\hat{V}\left(\varkappa\frac{\boldsymbol{\tau}_i}{(-\ln\tilde{H}(\boldsymbol{\tau}_i))}\right) + \theta(\boldsymbol{\tau}_i)\hat{W}\left(\varkappa\frac{\boldsymbol{\tau}_i}{(-\ln\tilde{H}(\boldsymbol{\tau}_i))}\right)\right) = \Theta\left(\varkappa\frac{\boldsymbol{\tau}_i}{(-\ln\tilde{H}(\boldsymbol{\tau}_i))}\right).$$

Finally, fix $\varkappa$ and deduce that

$$\sqrt{k_n}(\hat{\theta}_n^{(2)}(\boldsymbol{\tau}_i) - \theta(\boldsymbol{\tau}_i))_{i=1,\ldots,m} \Rightarrow \left(\Theta\left(\varkappa\frac{\boldsymbol{\tau}_i}{(-\ln\tilde{H}(\boldsymbol{\tau}_i))}\right)\right)_{i=1,\ldots,m}. \qquad \square$$

## 7. Discussion

In this paper, we have developed new estimators for the multivariate extremal index function. In order to construct scale invariant estimators, we have used a homogeneous transformation for the first estimator, but it leads to the question of the choice of the optimal transformation. We have also considered a second estimator which is scale invariant without transformation. One may also exploit averaging methods and consider, for example, the estimator defined by

$$\hat{\theta}_n^{(3)}(\boldsymbol{\tau}) = \frac{1}{\phi-\sigma}\int_\sigma^\phi \frac{-\ln\hat{H}_n(\varkappa\boldsymbol{\tau}_0)}{-\ln\widehat{\tilde{H}}_n(\varkappa\boldsymbol{\tau}_0)}\,d\varkappa, \qquad \boldsymbol{\tau}\in\mathbb{L}_{\boldsymbol{\tau}_0}\equiv\{\varkappa\boldsymbol{\tau}_0:\varkappa>0\},$$

where $0 < \sigma < \phi < \infty$. We have studied the weak convergence of our estimators as pointwise estimators and given their asymptotic distributions. To study their weak convergence as functional estimators, one must construct a specific functional space which is different from the Skorohod space of caglad functions which does not contain the set of scale invariant functions, then study their asymptotic properties in this space. This seems to be an important avenue for future research.

## Appendix A. Proofs of Propositions 2.1 and 4.1

**Proof of Proposition 2.1.** Let $\lambda > 0$. We have

$$
\begin{aligned}
&nP(Z_l^{(\boldsymbol{\tau})} > n\lambda^{-1}) \\
&= nP\left(\bigcup_{i=1}^{d}((1 - F_{i,-}(X_{l,i}))^{-1} > n(\tau_i\lambda)^{-1})\right) \\
&= nP\left(\bigcup_{i=1}^{d}(X_{l,i} > F_{i,-}^{\leftarrow}(1 - n^{-1}\tau_i\lambda))\right),
\end{aligned}
$$

where $F_{i,-}^{\leftarrow}(\tau) = \inf\{x \in \mathbb{R} : F_i(x) > \tau\}$. Since $F_i^{\leftarrow}(\tau) \leq F_{i,-}^{\leftarrow}(\tau)$ for each $\tau \in (0,1)$, we have

$$
\begin{aligned}
0 &\leq nP\left(\bigcup_{i=1}^{d}(X_{l,i} > u_{n,i}(\tau_i\lambda))\right) - nP\left(\bigcup_{i=1}^{d}(X_{l,i} > F_{i,-}^{\leftarrow}(1 - n^{-1}\tau_i\lambda))\right) \\
&\leq nP\left(\bigcup_{i=1}^{d}(F_i^{\leftarrow}(1 - n^{-1}\tau_i\lambda) < X_{l,i} \leq F_{i,-}^{\leftarrow}(1 - n^{-1}\tau_i\lambda))\right) \\
&= nP\left(\bigcup_{i=1}^{d}(X_{l,i} = F_{i,-}^{\leftarrow}(1 - n^{-1}\tau_i\lambda))\right) \\
&\leq \sum_{i=1}^{d} nP(X_{l,i} = F_{i,-}^{\leftarrow}(1 - n^{-1}\tau_i\lambda)).
\end{aligned}
$$

Note that

$$
\lim_{n \to \infty} \frac{P(X_{l,i} = F_{i,-}^{\leftarrow}(1 - n^{-1}\tau_i\lambda))}{\tau_i\lambda/n} = \lim_{n \to \infty} \frac{P(X_{l,i} = F_{i,-}^{\leftarrow}(1 - n^{-1}\tau_i\lambda))}{P(X_{l,i} > u_{n,i}(\tau_i\lambda))}
$$

$$
= \lim_{x \to x_{f,i}} \frac{\bar{F}_i(x) - \bar{F}_i(x-)}{\bar{F}_i(x)} = 0
$$

and then

$$\lim_{n\to\infty} nP(Z_l^{(\boldsymbol{\tau})} > n\lambda^{-1}) = \lim_{n\to\infty} nP\left(\bigcup_{i=1}^{d}(X_{l,i} > u_{n,i}(\tau_i\lambda))\right).$$

By (1.2) and the homogeneity property of $-\ln\tilde{H}$, it follows that

$$\lim_{n\to\infty} nP(Z_l^{(\boldsymbol{\tau})} > n\lambda^{-1}) = \lim_{n\to\infty} n(1 - F(\boldsymbol{u}_n(\lambda\boldsymbol{\tau}))) = -\ln\tilde{H}(\lambda\boldsymbol{\tau}) = \lambda(-\ln\tilde{H}(\boldsymbol{\tau})).$$

By taking $\lambda = \varkappa(-\ln\tilde{H}(\boldsymbol{\tau}))^{-1}$, we deduce that

$$\lim_{n\to\infty} nP(Z_l^{(\boldsymbol{\tau})} > v_n^{(\boldsymbol{\tau})}(\varkappa)) = \varkappa.$$

We now have

$$P\left(\max_{l=1,\ldots,n} Z_l^{(\boldsymbol{\tau})} \leq v_n^{(\boldsymbol{\tau})}(\varkappa)\right)$$

$$= P\left(\max_{l=1,\ldots,n}\max_{i=1,\ldots,d} \tau_i Y_{l,i} \leq n(-\ln\tilde{H}(\boldsymbol{\tau}))\varkappa^{-1}\right)$$

$$= P\left(\max_{i=1,\ldots,d} \tau_i \max_{l=1,\ldots,n} Y_{l,i} \leq n(-\ln\tilde{H}(\boldsymbol{\tau}))\varkappa^{-1}\right)$$

$$= P\left(\max_{i=1,\ldots,d} \frac{\tau_i}{1 - F_{i,-}(M_{n,i})} \leq n(-\ln\tilde{H}(\boldsymbol{\tau}))\varkappa^{-1}\right)$$

$$= P(M_{n,i} \leq F_{i,-}^{\leftarrow}(1 - n^{-1}\varkappa\tau_i(-\ln\tilde{H}(\boldsymbol{\tau}))^{-1}), i=1,\ldots,d).$$

Note that

$$P(M_{n,i} \leq F_{i,-}^{\leftarrow}(1 - n^{-1}\varkappa\tau_i(-\ln\tilde{H}(\boldsymbol{\tau}))^{-1}), i=1,\ldots,d)$$
$$- P(M_{n,i} \leq F_i^{\leftarrow}(1 - n^{-1}\varkappa\tau_i(-\ln\tilde{H}(\boldsymbol{\tau}))^{-1}), i=1,\ldots,d)$$
$$= P(M_{n,i} = F_{i,-}^{\leftarrow}(1 - n^{-1}\varkappa\tau_i(-\ln\tilde{H}(\boldsymbol{\tau}))^{-1}), i=1,\ldots,d)$$
$$\leq \sum_{i=1}^{d} nP(X_{l,i} = F_{i,-}^{\leftarrow}(1 - n^{-1}\varkappa\tau_i(-\ln\tilde{H}(\boldsymbol{\tau}))^{-1})) \xrightarrow[n\to\infty]{} 0.$$

It follows that

$$\lim_{n\to\infty} P(M_n^{(Z)} \leq v_n^{(\boldsymbol{\tau})}(\varkappa))$$
$$= \lim_{n\to\infty} P(M_{n,i} \leq F_i^{\leftarrow}(1 - n^{-1}\varkappa\tau_i(-\ln\tilde{H}(\boldsymbol{\tau}))^{-1}), i=1,\ldots,d)$$
$$= H(\varkappa\boldsymbol{\tau}(-\ln\tilde{H}(\boldsymbol{\tau}))^{-1}) = \tilde{H}(\varkappa\boldsymbol{\tau}(-\ln\tilde{H}(\boldsymbol{\tau}))^{-1})^{\theta(\varkappa\boldsymbol{\tau}(-\ln\tilde{H}(\boldsymbol{\tau}))^{-1})}$$
$$= e^{-\theta(\boldsymbol{\tau})\varkappa},$$

which means that $\theta(\boldsymbol{\tau})$ is the univariate extremal index of the stationary sequence $(Z_n^{(\boldsymbol{\tau})})_{n\geq 1}$. □

**Proof of Proposition 4.1.** The proof follows the corresponding lines of the proof of Proposition 1 in [30]. Let $\boldsymbol{\tau}_1, \boldsymbol{\tau}_2 \in [0,\infty)^d \setminus \{\boldsymbol{0}\}$ and $s$ be a positive constant. Define

$$R_n(sn, \boldsymbol{\tau}_1, \boldsymbol{\tau}_2) = (N_{n,0,sn}^{(\boldsymbol{\tau}_1, \boldsymbol{\tau}_2)}, N_{n,1,sn}^{(\boldsymbol{\tau}_1, \boldsymbol{\tau}_2)}, N_{n,2,sn}^{(\boldsymbol{\tau}_1, \boldsymbol{\tau}_2)}, N_{n,3,sn}^{(\boldsymbol{\tau}_1, \boldsymbol{\tau}_2)})'.$$

By considering a $\Delta(\boldsymbol{u}_n(\boldsymbol{\tau}_1), \boldsymbol{u}_n(\boldsymbol{\tau}_2))$-separating sequence, $(r_n)_{n\geq 1}$, and Berstein's blocks method (see Lemma 2.2 in [15] or the proof of Lemma 6.7 in [37]), we get

$$\lim_{n\to\infty} |\mathbb{E}(e^{iv' R_n(sn,\boldsymbol{\tau}_1,\boldsymbol{\tau}_2)}) - (\mathbb{E}(e^{iv' R_n(r_n,\boldsymbol{\tau}_1,\boldsymbol{\tau}_2)}))^{m_n}| = 0,$$

where $m_n = \lfloor sn/r_n \rfloor$, $\lfloor x \rfloor$ denotes the integer part of $x$ and $v \in \mathbb{R}^4$. Now, note that

$$\mathbb{E}(e^{iv' R_n(r_n,\boldsymbol{\tau}_1,\boldsymbol{\tau}_2)}) = P(N_{n,0,r_n}^{(\boldsymbol{\tau}_1,\boldsymbol{\tau}_2)} = 0) + \mathbb{E}(e^{iv' R_n(r_n,\boldsymbol{\tau}_1,\boldsymbol{\tau}_2)} | N_{n,0,r_n}^{(\boldsymbol{\tau}_1,\boldsymbol{\tau}_2)} > 0) P(N_{n,0,r_n}^{(\boldsymbol{\tau}_1,\boldsymbol{\tau}_2)} > 0).$$

Since $(r_n)_{n\geq 1}$ is a $\Delta(\boldsymbol{u}_n(\boldsymbol{\tau}_1), \boldsymbol{u}_n(\boldsymbol{\tau}_2))$-separating sequence and

$$\lim_{n\to\infty} P(N_{n,0,n}^{(\boldsymbol{\tau}_1,\boldsymbol{\tau}_2)} = 0) = e^{-\theta(\boldsymbol{\tau}_1 \vee \boldsymbol{\tau}_2) \ln(\tilde{H}(\boldsymbol{\tau}_1 \vee \boldsymbol{\tau}_2))},$$

we have

$$\lim_{n\to\infty} \frac{n}{r_n} P(N_{n,0,r_n}^{(\boldsymbol{\tau}_1,\boldsymbol{\tau}_2)} > 0) = -\theta(\boldsymbol{\tau}_1 \vee \boldsymbol{\tau}_2) \ln(\tilde{H}(\boldsymbol{\tau}_1 \vee \boldsymbol{\tau}_2)).$$

We then deduce that

$$\mathbb{E}(e^{iv' R_n(sn,\boldsymbol{\tau}_1,\boldsymbol{\tau}_2)}) = \exp(m_n P(N_{n,0,r_n}^{(\boldsymbol{\tau}_1,\boldsymbol{\tau}_2)} > 0) \mathbb{E}(e^{iv' R_n(r_n,\boldsymbol{\tau}_1,\boldsymbol{\tau}_2)} - 1 | N_{n,0,r_n}^{(\boldsymbol{\tau}_1,\boldsymbol{\tau}_2)} > 0)) + o(1).$$

On one hand, we have

$$\lim_{n\to\infty} \mathbb{E}(e^{iv' R_n(sn,\boldsymbol{\tau}_1,\boldsymbol{\tau}_2)}) = \exp(-s\theta(\boldsymbol{\tau}_1 \vee \boldsymbol{\tau}_2) \ln(\tilde{H}(\boldsymbol{\tau}_1 \vee \boldsymbol{\tau}_2))(\mathbb{E}e^{iv' \zeta_l^{(\boldsymbol{\tau}_1,\boldsymbol{\tau}_2)}} - 1))$$

where $\zeta_l^{(\boldsymbol{\tau}_1,\boldsymbol{\tau}_2)} = (\zeta_{1,l}^{(\boldsymbol{\tau}_1,\boldsymbol{\tau}_2)} + \zeta_{2,l}^{(\boldsymbol{\tau}_1,\boldsymbol{\tau}_2)} + \zeta_{3,l}^{(\boldsymbol{\tau}_1,\boldsymbol{\tau}_2)}, \zeta_{1,l}^{(\boldsymbol{\tau}_1,\boldsymbol{\tau}_2)}, \zeta_{2,l}^{(\boldsymbol{\tau}_1,\boldsymbol{\tau}_2)}, \zeta_{3,l}^{(\boldsymbol{\tau}_1,\boldsymbol{\tau}_2)})'$. In particular, we derive the weak convergence of the sequence $(R_n(n, \boldsymbol{\tau}_1, \boldsymbol{\tau}_2))_{n\geq 1}$ by choosing $s = 1$.

On the other hand, it is easily seen by using the definition of $\boldsymbol{u}_n(\boldsymbol{\tau})$ that

$$\lim_{n\to\infty} \mathbb{E}(e^{iv' R_n(sn,\boldsymbol{\tau}_1,\boldsymbol{\tau}_2)}) = \lim_{n\to\infty} \mathbb{E}(e^{iv' R_{sn}(sn,s\boldsymbol{\tau}_1,s\boldsymbol{\tau}_2)}) = \lim_{n\to\infty} \mathbb{E}(e^{iv' R_n(n,s\boldsymbol{\tau}_1,s\boldsymbol{\tau}_2)})$$

$$= \exp(-s\theta(\boldsymbol{\tau}_1 \vee \boldsymbol{\tau}_2) \ln(\tilde{H}(\boldsymbol{\tau}_1 \vee \boldsymbol{\tau}_2))(\mathbb{E}e^{iv' \zeta^{(s\boldsymbol{\tau}_1,s\boldsymbol{\tau}_2)}} - 1)).$$

Therefore, $\mathbb{E}e^{iv' \zeta^{(s\boldsymbol{\tau}_1,s\boldsymbol{\tau}_2)}} = \mathbb{E}e^{iv' \zeta^{(\boldsymbol{\tau}_1,\boldsymbol{\tau}_2)}}$ and it follows that $\pi^{(\boldsymbol{\tau}_1,\boldsymbol{\tau}_2)}$ is scale invariant. □

## Appendix B. Covariance function of $\Theta$

Let us define the functions $C_{i,j}(\cdot,\cdot)$ for $i = 1, 2$ and $j = 1, 2$ by

$$C_{1,1}(\boldsymbol{\tau}_1, \boldsymbol{\tau}_2) = H(\boldsymbol{\tau}_1 \vee \boldsymbol{\tau}_2) - H(\boldsymbol{\tau}_1)H(\boldsymbol{\tau}_2),$$

$$C_{2,2}(\boldsymbol{\tau}_1, \boldsymbol{\tau}_2) = -\theta(\boldsymbol{\tau}_1 \vee \boldsymbol{\tau}_2) \ln(\tilde{H}(\boldsymbol{\tau}_1 \vee \boldsymbol{\tau}_2)) \mathbb{E}((\zeta_1^{(\boldsymbol{\tau}_1,\boldsymbol{\tau}_2)} + \zeta_3^{(\boldsymbol{\tau}_1,\boldsymbol{\tau}_2)})(\zeta_2^{(\boldsymbol{\tau}_1,\boldsymbol{\tau}_2)} + \zeta_3^{(\boldsymbol{\tau}_1,\boldsymbol{\tau}_2)})),$$

$$C_{1,2}(\boldsymbol{\tau}_1, \boldsymbol{\tau}_2) = \mathbb{E}(N_2^{(\boldsymbol{\tau}_1,\boldsymbol{\tau}_2)} 1_{\{N_1^{(\boldsymbol{\tau}_1,\boldsymbol{\tau}_2)}=0, N_3^{(\boldsymbol{\tau}_1,\boldsymbol{\tau}_2)}=0\}}) + H(\boldsymbol{\tau}_1) \ln \tilde{H}(\boldsymbol{\tau}_2),$$

$$C_{2,1}(\boldsymbol{\tau}_1, \boldsymbol{\tau}_2) = \mathbb{E}(N_1^{(\boldsymbol{\tau}_1,\boldsymbol{\tau}_2)} 1_{\{N_2^{(\boldsymbol{\tau}_1,\boldsymbol{\tau}_2)}=0, N_3^{(\boldsymbol{\tau}_1,\boldsymbol{\tau}_2)}=0\}}) + H(\boldsymbol{\tau}_2) \ln \tilde{H}(\boldsymbol{\tau}_1).$$

Note that $C_{1,2}(\boldsymbol{\tau}_2, \boldsymbol{\tau}_1) = C_{2,1}(\boldsymbol{\tau}_1, \boldsymbol{\tau}_2)$.

Let us now characterize the covariance function of the process $\Theta$. We have

$$\begin{aligned}
&\text{cov}(\Theta(\boldsymbol{\tau}_1), \Theta(\boldsymbol{\tau}_2)) \\
&= \frac{1}{H(\boldsymbol{\tau}_1) \ln \tilde{H}(\boldsymbol{\tau}_1)} \frac{1}{H(\boldsymbol{\tau}_2) \ln \tilde{H}(\boldsymbol{\tau}_2)} G_{1,1}(\boldsymbol{\tau}_1, \boldsymbol{\tau}_2) + \frac{\theta(\boldsymbol{\tau}_2)}{H(\boldsymbol{\tau}_1) \ln \tilde{H}(\boldsymbol{\tau}_1)} G_{1,2}(\boldsymbol{\tau}_1, \boldsymbol{\tau}_2) \\
&\quad + \frac{\theta(\boldsymbol{\tau}_1)}{H(\boldsymbol{\tau}_2) \ln \tilde{H}(\boldsymbol{\tau}_2)} G_{2,1}(\boldsymbol{\tau}_1, \boldsymbol{\tau}_2) + \theta(\boldsymbol{\tau}_1)\theta(\boldsymbol{\tau}_2) G_{2,2}(\boldsymbol{\tau}_1, \boldsymbol{\tau}_2),
\end{aligned}$$

where

$$\begin{aligned}
G_{1,1}(\boldsymbol{\tau}_1, \boldsymbol{\tau}_2) &= C_{1,1}(\boldsymbol{\tau}_1, \boldsymbol{\tau}_2) - H(\boldsymbol{\tau}_1) \sum_{i=1}^{d} \frac{\partial \ln H(\boldsymbol{\tau}_1)}{\partial \tau_{i,1}} C_{2,1}(\boldsymbol{\tau}_1^{(i)}, \boldsymbol{\tau}_2) \\
&\quad - H(\boldsymbol{\tau}_2) \sum_{i=1}^{d} \frac{\partial \ln H(\boldsymbol{\tau}_2)}{\partial \tau_{i,2}} C_{1,2}(\boldsymbol{\tau}_1, \boldsymbol{\tau}_2^{(i)}) \\
&\quad + H(\boldsymbol{\tau}_1) H(\boldsymbol{\tau}_2) \sum_{i=1}^{d} \sum_{j=1}^{d} \frac{\partial \ln H(\boldsymbol{\tau}_1)}{\partial \tau_{i,1}} \frac{\partial \ln H(\boldsymbol{\tau}_2)}{\partial \tau_{j,2}} C_{2,2}(\boldsymbol{\tau}_1^{(i)}, \boldsymbol{\tau}_2^{(j)}),
\end{aligned}$$

$$\begin{aligned}
G_{1,2}(\boldsymbol{\tau}_1, \boldsymbol{\tau}_2) &= C_{1,2}(\boldsymbol{\tau}_1, \boldsymbol{\tau}_2) - H(\boldsymbol{\tau}_1) \sum_{i=1}^{d} \frac{\partial \ln H(\boldsymbol{\tau}_1)}{\partial \tau_{i,1}} C_{2,2}(\boldsymbol{\tau}_1^{(i)}, \boldsymbol{\tau}_2) \\
&\quad + \sum_{i=1}^{d} \frac{\partial \ln \tilde{H}(\boldsymbol{\tau}_2)}{\partial \tau_{i,2}} C_{1,2}(\boldsymbol{\tau}_1, \boldsymbol{\tau}_2^{(i)}) \\
&\quad - H(\boldsymbol{\tau}_1) \sum_{i=1}^{d} \sum_{j=1}^{d} \frac{\partial \ln H(\boldsymbol{\tau}_1)}{\partial \tau_{i,1}} \frac{\partial \ln \tilde{H}(\boldsymbol{\tau}_2)}{\partial \tau_{j,2}} C_{2,2}(\boldsymbol{\tau}_1^{(i)}, \boldsymbol{\tau}_2^{(j)}),
\end{aligned}$$

$$G_{2,1}(\boldsymbol{\tau}_1, \boldsymbol{\tau}_2) = C_{2,1}(\boldsymbol{\tau}_1, \boldsymbol{\tau}_2) + \sum_{i=1}^{d} \frac{\partial \ln \tilde{H}(\boldsymbol{\tau}_1)}{\partial \tau_{i,1}} C_{2,1}(\boldsymbol{\tau}_1^{(i)}, \boldsymbol{\tau}_2)$$

$$- H(\boldsymbol{\tau}_2) \sum_{i=1}^{d} \frac{\partial \ln H(\boldsymbol{\tau}_2)}{\partial \tau_{i,2}} C_{2,2}(\boldsymbol{\tau}_1, \boldsymbol{\tau}_2^{(i)})$$

$$- H(\boldsymbol{\tau}_2) \sum_{i=1}^{d} \sum_{j=1}^{d} \frac{\partial \ln \tilde{H}(\boldsymbol{\tau}_1)}{\partial \tau_{i,1}} \frac{\partial \ln H(\boldsymbol{\tau}_2)}{\partial \tau_{j,2}} C_{2,2}(\boldsymbol{\tau}_1^{(i)}, \boldsymbol{\tau}_2^{(j)}),$$

$$G_{2,2}(\boldsymbol{\tau}_1, \boldsymbol{\tau}_2) = C_{2,2}(\boldsymbol{\tau}_1, \boldsymbol{\tau}_2) + \sum_{i=1}^{d} \frac{\partial \ln \tilde{H}(\boldsymbol{\tau}_1)}{\partial \tau_{i,1}} C_{2,2}(\boldsymbol{\tau}_1^{(i)}, \boldsymbol{\tau}_2)$$

$$+ \sum_{i=1}^{d} \frac{\partial \ln \tilde{H}(\boldsymbol{\tau}_2)}{\partial \tau_{i,2}} C_{2,2}(\boldsymbol{\tau}_1, \boldsymbol{\tau}_2^{(i)})$$

$$+ \sum_{i=1}^{d} \sum_{j=1}^{d} \frac{\partial \ln \tilde{H}(\boldsymbol{\tau}_1)}{\partial \tau_{i,1}} \frac{\partial \ln \tilde{H}(\boldsymbol{\tau}_2)}{\partial \tau_{j,2}} C_{2,2}(\boldsymbol{\tau}_1^{(i)}, \boldsymbol{\tau}_2^{(j)}).$$

## Acknowledgements

The author gratefully acknowledges the anonymous referees for various suggestions leading to improved results throughout the paper.